\documentclass[twocolumn]{aastex61}

\received{2018 March 1}
\accepted{2018 May 17}
\submitjournal{PASP}

\shorttitle{Variability analysis of $\delta$\,Scuti candidate stars}
\shortauthors{Pak\v{s}tien\.{e} et al.}

\begin{document}

\title{Variability analysis of $\delta$\,Scuti candidate stars}

\correspondingauthor{Erika Pak\v stien\.e}
\email{erika.pakstiene@tfai.vu.lt}

\author{Erika Pak\v stien\.e}
\affil{Institute of Theoretical Physics and Astronomy, Vilnius University, Saul\.{e}tekio av. 3, 10257 Vilnius, Lithuania}

\author{Rimvydas Janulis}
\affil{Institute of Theoretical Physics and Astronomy, Vilnius University, Saul\.{e}tekio av. 3, 10257 Vilnius, Lithuania}

\author{Gra\v zina Tautvai\v{s}ien\.{e}}
\affil{Institute of Theoretical Physics and Astronomy, Vilnius University, Saul\.{e}tekio av. 3, 10257 Vilnius, Lithuania}

\author{Arnas Drazdauskas}
\affil{Institute of Theoretical Physics and Astronomy, Vilnius University, Saul\.{e}tekio av. 3, 10257 Vilnius, Lithuania}

\author{Lukas Klebonas}
\affil{Institute of Theoretical Physics and Astronomy, Vilnius University, Saul\.{e}tekio av. 3, 10257 Vilnius, Lithuania}
\affil{Mathematisch-Naturwissenschaftliche Fakult\"at, Universit\"at Bonn, Wegelerstra{\ss}e 10, D-53115 Bonn, Germany}

\author{ \v{S}ar\= unas Mikolaitis}
\affil{Institute of Theoretical Physics and Astronomy, Vilnius University, Saul\.{e}tekio av. 3, 10257 Vilnius, Lithuania}

\author{Renata Minkevi\v{c}i\={u}t\.e}
\affil{Institute of Theoretical Physics and Astronomy, Vilnius University, Saul\.{e}tekio av. 3, 10257 Vilnius, Lithuania}

\author{Vilius Bagdonas}
\affil{Institute of Theoretical Physics and Astronomy, Vilnius University, Saul\.{e}tekio av. 3, 10257 Vilnius, Lithuania}



\begin{abstract}

The Hipparcos catalog contains stars suspected to be $\delta$\,Scuti variables for which extensive ground-based observations and characterization of variability are necessary. Our aim was to characterize variability of 13 candidates to $\delta$\,Scuti type stars. We obtained 24\,215 CCD images and analyzed stellar light curves using the Period04 program.  
Twelve $\delta$\,Scuti candidate stars have been characterized as pulsating with frequencies intrinsic to  $\delta$\,Scuti stars: HIP\,2923, HIP\,5526, HIP\,5659, HIP\,11090, HIP\,17585, HIP\,74155, HIP\,101473, HIP\,106219,  HIP\,107786, HIP\,113487, HIP\,115093, HIP\,115856. Five of them (HIP\,2923, HIP\,5526, HIP\,11090, HIP\,115856 and HIP\,106219) may be hybrid $\delta$\,Scuti-$\gamma$\,Doradus pulsators.
One more candidate, HIP 106223, is a variable star with longer periods of pulsations which are intrinsic to $\gamma$\,Doradus.

\end{abstract}


\section{Introduction} \label{sec:intro}

The $\delta$\,Scuti type stars are pulsating stars situated in the classical Cepheid instability strip \citep{Breger2000}. Most of the $\delta$\,Scuti pulsators are stars of spectral types A0$-$F5\,III$-$V. They are main-sequence or immediate
post-main-sequence variable stars moving to the giant branch \citep{Breger2000}. 
Pulsation periods of $\delta$\,Scuti type stars vary between $~$0.008 and 0.42~days \citep{Sanchez2017}. Excited modes have amplitudes from 0.001~mag \citep{Breger2000} up to almost one magnitude in blue bands \citep{Sanchez2017}.

Majority of $\delta$\,Scuti type stars have multiple non-radial pulsation modes, while some of them are pure radial pulsators \citep{Breger2000}. A nature of the excited modes can be complicated: either pure $p$-modes, or pure $g$- or mixed $p$- and $g$-modes.  Oscillations of $\delta$\,Scuti stars are not fully understood. There are many modes which are  theoretically expected to be excited in a given frequency range but not all modes in this range are detected \citep{Goupil2005}.

$\gamma$\,Doradus stars is another group of variable stars existing in close neighbourhood with $\delta$\,Scuti stars. Some stars are hybrid $\delta$\,Scuti-$\gamma$\,Doradus pulsators showing high-frequency $p$-mode pulsations typical of $\delta$\,Scuti stars and low-frequency $g$-mode oscillations characteristic of $\gamma$\,Doradus stars. 
Only high amplitude low frequency variations of $\gamma$\,Doradus and hybrid stars may be observed from the ground. Low amplitude frequencies can be detected only from space telescope observations. Observations from space missions such as MOST, CoRoT, and $Kepler$ have revealed a large number of hybrid $\delta$\,Scuti-$\gamma$\,Doradus pulsators, which are situated where the instability strips of $\delta$\,Scuti and $\gamma$\,Doradus stars partially overlap in the Hertzpsrung--Russell (HR) diagram. These stars show behavior typical to hybrid $\delta$\,Scuti--$\gamma$\,Doradus pulsators (\citealt{Grigahcene2010}, \citealt{Uytterhoeven2011}, \citealt{Bradley2015}, \citealt{Xiong2016}, \citealt{Sanchez2017}).

A difficult task is to identify the modes. The method of mode equidistances is not always working as frequencies of some modes do not follow the pure rules. A direct fit of theoretical models to observed frequencies is
difficult as often several choices of stellar models are possible within uncertainties. Pulsation modes are very sensitive to the convection treatment, as a reliable description of time dependent convection
is necessary. A fast rotation of this type of stars makes additional problems in the mode identification framework \citep{Goupil2005}.

Though the group of $\delta$\,Scuti type stars is among most numerous groups of pulsators, it is also one of the least understood groups of stars. Much more observational information about this type of stars is necessary in order to improve their models and uncover details about processes happening beneath their surfaces. 

\section{Selected targets}

\begin{figure}[!ht]
   \centering
   \includegraphics[width=\hsize]{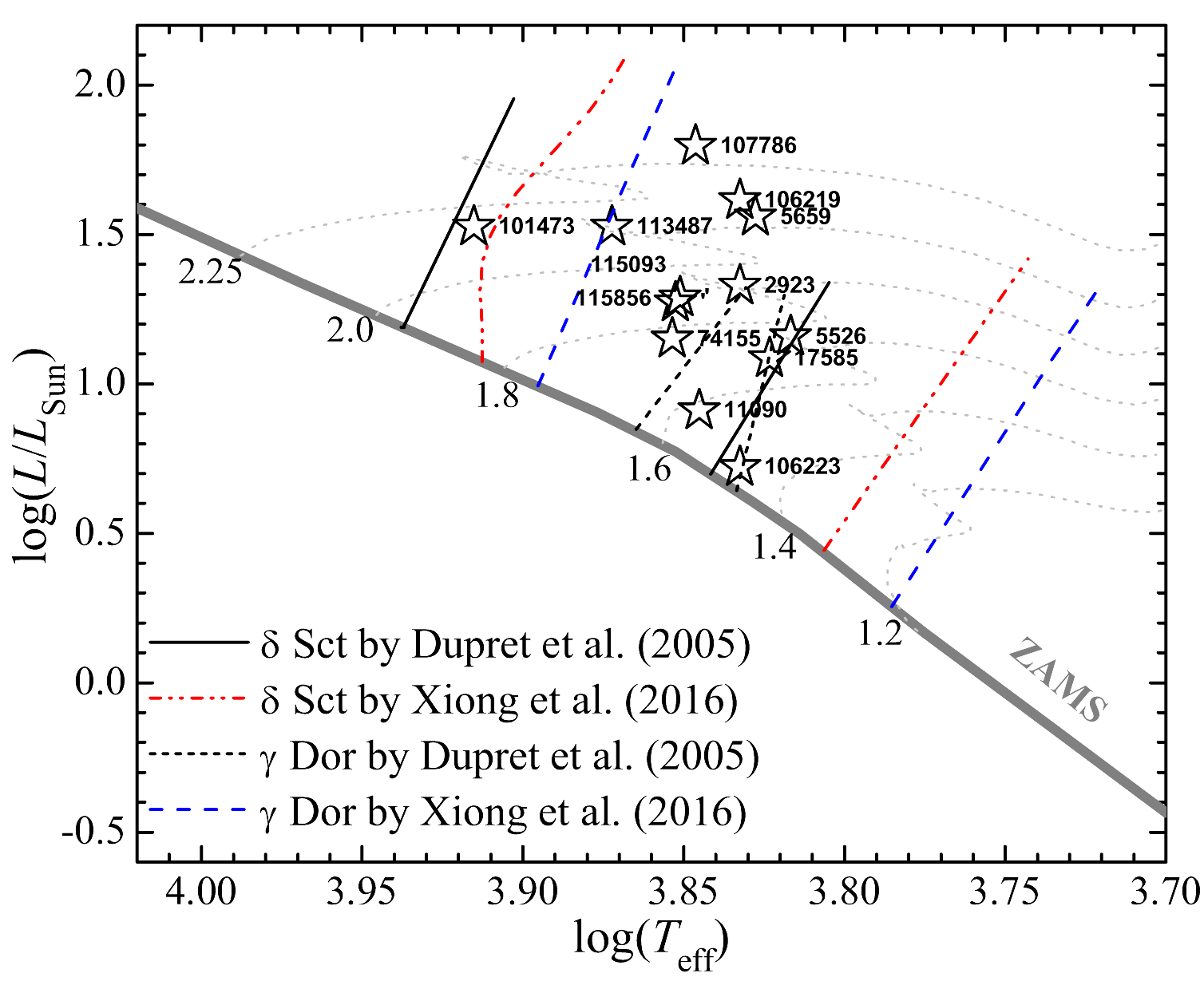}
      \caption{Positions of analyzed $\delta$\,Scuti candidates and instability strips. See text for more explanations.} 
         \label{Fig1_strip}
   \end{figure}

We chose for observations the $\delta$\,Scuti candidates selected from the Hipparcos catalog by \cite{Handler2002} and selected 13 stars suitable for observations with telescopes of the Mol\.etai Astronomical Observatory (MAO) in Lithuania. 
We list the selected $\delta$\,Scuti candidates in Table~\ref{table:1} and show their positions in the HR diagram (Figure~\ref{Fig1_strip}). $T_{\rm eff}$ and $L/L_{\rm Sun}$ were taken from \cite{McDonald2012}. Positions of theoretical instability strips for $\delta$ Scuti and $\gamma$\,Doradus were taken from \cite{Dupret2005} and \cite{Xiong2016}.

\begin{table*}
\caption{Information about observed stars. }             
\label{table:1}      
\begin{tabular}{l c c c  c c c c c c}        
\hline               
Star     &  $\alpha$(2000)   & $\delta$(2000)   & $V$  	 &Sp.type& Images      & Runs &Point error&	STD of LC &Comparison 	\\
         &      h m s        &$^\circ$  $^\prime$ $^{\prime\prime}$ & mag &(Simbad) & &  & (mean)  &             &star \\
\hline 
HIP 2923    & 00 37 03.56  &  +31 29 11.31  &  7.65 &	F0III	  &2965   &17	  &13.50      	&22.76     &TYC 2275-1038-1 \\	
HIP 5526    & 01 10 43.31  &  +27 52 04.61  &  8.10&	F0	  &1684	  &12	  &13.24      	&33.20     &TYC 1753-1926-1\\	
HIP 5659    & 01 12 41.26  &  +65 00 32.83  &  7.62&	F0	  &1556	  &12	  &8.30	      	&23.04     &TYC 4038-716-1\\	
HIP 11090   & 02 22 50.30  &  +41 23 46.67  &  5.80&	F0III-IV  &1682	  &12	  &6.16	      	&21.70     &TYC 2835-85-1\\		
HIP 17585   & 03 46 00.94  &  +67 12 05.78  &  5.79&	F0IV	  &3475	  &14	  &14.82      	&29.91     &TYC 4075-932-1\\	
HIP 74155   & 15 09 06.24  &  +69 39 11.11  &  7.12&	F2	  &175	  &6	  &10.88      	&19.33     &TYC 4411-835-1\\	
HIP 101473  & 20 33 53.70  &  +10 03 35.05  &  6.54&	A2Vnn	  &572	  &7	  &6.30	      	&15.75     &TYC 1092-1447-1\\	
HIP 106219  & 21 30 53.28  &  +24 46 51.90  &  8.25&	A5	  &1312	  &14	  &5.41	      	&12.12     &TYC 2192-608-1\\	
HIP 106223  & 21 30 57.05  &  +16 34 15.57  &  8.64&	A5 	  &3531	  &28	  &15.62      	&43.59     &TYC 1664-433-1\\	
HIP 107786  & 21 50 08.23  &  +19 25 26.38  &  7.21&	A5	  &1215	  &14	  &10.38      	&22.39     &TYC 1674-299-1\\	
HIP 113487  & 22 58 58.96  &  +34 04 00.67  &  7.54&	A0	  &916	  &11	  &8.06	      	&29.04     &TYC 2758-490-1\\	
HIP 115093  & 23 18 42.26  &  +36 05 24.81  &  7.37&	F0	  &1390	  &10	  &8.10	      	&16.69     &TYC 2764-1570-1\\ 	
HIP 115856  & 23 28 23.54  &  +19 53 08.09  &  6.67&	F0	  &3742	  &17	  &10.22      	&26.84     &TYC 1726-1519-1\\
\hline                                   
\end{tabular}
\end{table*}

\cite{Dupret2005} calculated an instability strips of $\delta$\,Scuti and $\gamma$\,Doradus stars. These instability strips fit well ground-based observations. \cite{Dupret2005} also predicted hybrid stars in the overlapping
region of the $\delta$\,Scuti and $\gamma$\,Doradus instability strips.
 
Figure~\ref{Fig1_strip} shows also the more recently calculated instability strips of $\delta$\,Scuti and $\gamma$\,Doradus stars by \cite{Xiong2016}, which fit well the data collected by space telescopes. 
The Padova evolutionary tracks\footnote{http://pleiadi.pd.astro.it/} of six stars with different masses (thin dotted lines) and the Zero Age Main Sequence (ZAMS, thick gray line) are also shown in Figure~\ref{Fig1_strip}. 

Twelve of our selected $\delta$\,Scuti candidate stars obviously lay inside the instability strip of $\delta$\,Scuti or $\gamma$\,Doradus stars, and only one of the candidates is located close to the blue edge of $\delta$\,Scuti instability strip. Thus, we may expect to observe both types of oscillations in at least some targets.  

In Table~\ref{table:1} we present a list of targeted $\delta$\,Scuti candidates, their coordinates,  $V$ magnitudes, spectral types, a number of images taken, a number of runs, names of comparison stars used for the data reduction and parameters of data quality (mean errors of the observed points and standard deviations of the light curves). All the stars belong to the preliminary PLATO fields (\citealt{Miglio2017}): HIP\,5659, HIP\,17585, and HIP\,74155 belong to STEP02; HIP\,2923, HIP\,5526, and HIP\,74155 belong to STEP07; and the remaining stars belong to the field STEP05.  

\section{Observations}

Observations were performed with a 51~cm Maksutov-type MAO telescope of 35~cm working diameter of primary mirror and the Apogee Alta U47 CCD camera. 
This instrumentation allows us to observe bright stars without saturation of CCD pixels. 
For the observations we used the $Y$ filter of a medium-band Vilnius photometric system.
Its effective wavelength is at 466~nm and the width is 26~nm \citep{Straizys&Sviderskiene1972} which is close to the Johnson's $B$ filter, but is transparent for a narrower range of wavelengths. 

 The observations were carried out in a semi-robotic mode, i.e. the telescope was changing the pointing and took exposures of different fields of the sky according to the beforehand prepared script. This mode allowed us to observe light curves (LC) of stars in $5-7$ different fields of the sky during the same night with a cadence of $15-30$\,minutes. 
Observations were carried out using blind tracking without autoguiding, 
thus we calibrated the CCD images very carefully in order to avoid artificial signals.
We were taking more than 10 images of the bias, dark and sky flat fields during each night for the CCD image calibration.

\begin{figure}
   \centering
   \includegraphics[width=\hsize]{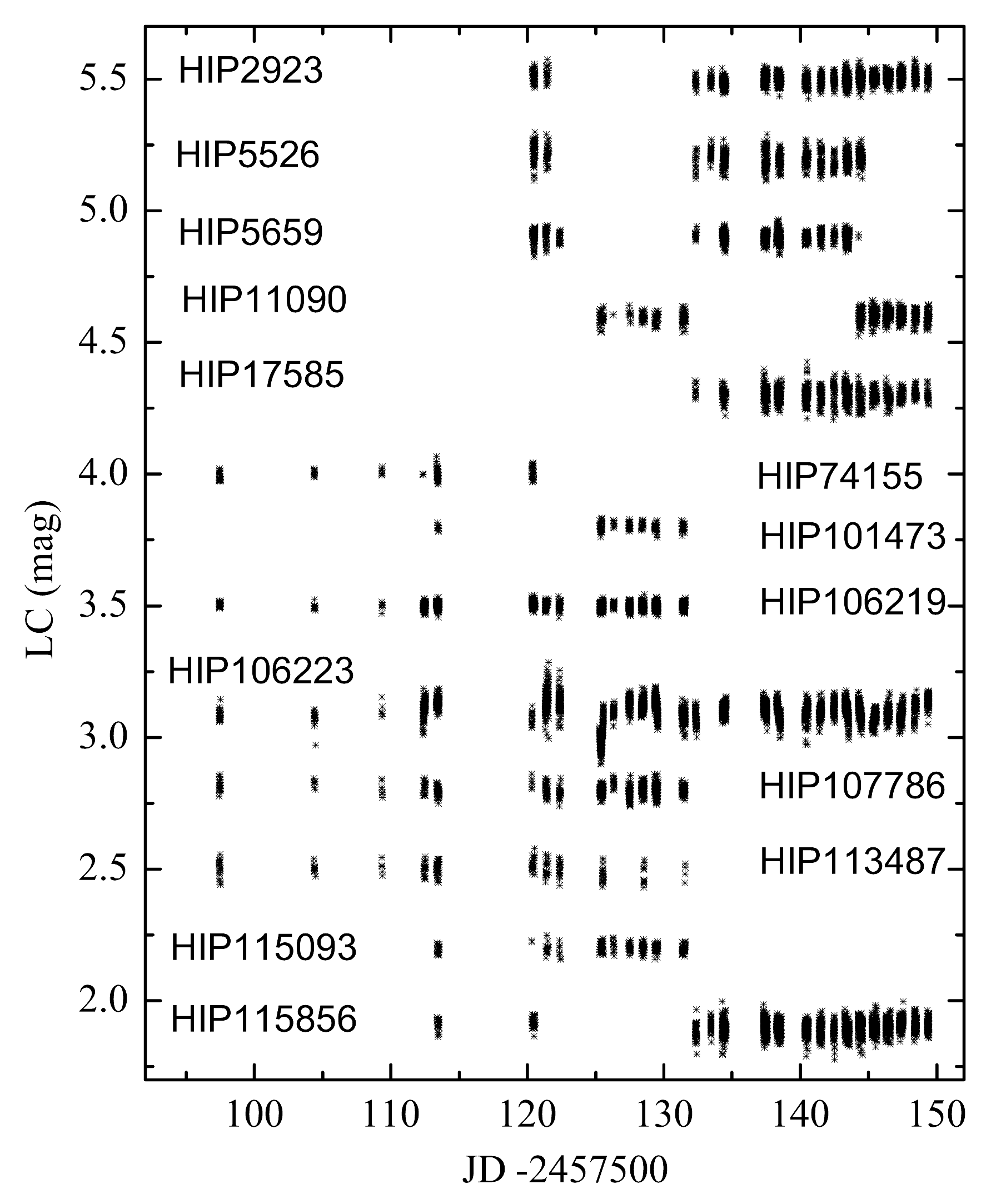}
      \caption{A layout of light curves for the investigated $\delta$\,Scuti candidate stars. }  
         \label{Fig2_all_LCs}
   \end{figure}

A layout of obtained LCs of the $\delta$\,Scuti candidates is presented in Figure~\ref{Fig2_all_LCs}. 

\section{Data reduction and analysis}

The observed images were first processed with the Muniwin program of the software package C-Munipack\footnote{http://c-munipack.sourceforge.net/}  \citep{Muniwin14}, which is built on the basis of a software package DAOPHOT for doing stellar photometry in crowded stellar fields \citep{Daophot87}. 
The Muniwin program is designed for the time series differential aperture photometry and searching of variable stars. 
We used the {\it Advanced} image calibration procedure, to perform the bias and dark frame subtraction, and flat-field correction.

We performed photometry with different apertures in order to select the best one, corresponding to a smallest standard deviation of the obtained light curves. The selected aperture was 4 or 5 pixels for a field. We used it to determine the instrumental magnitudes of all detected stars in the field. 

\begin{figure}[!ht]
   \centering
   \includegraphics[width=\hsize]{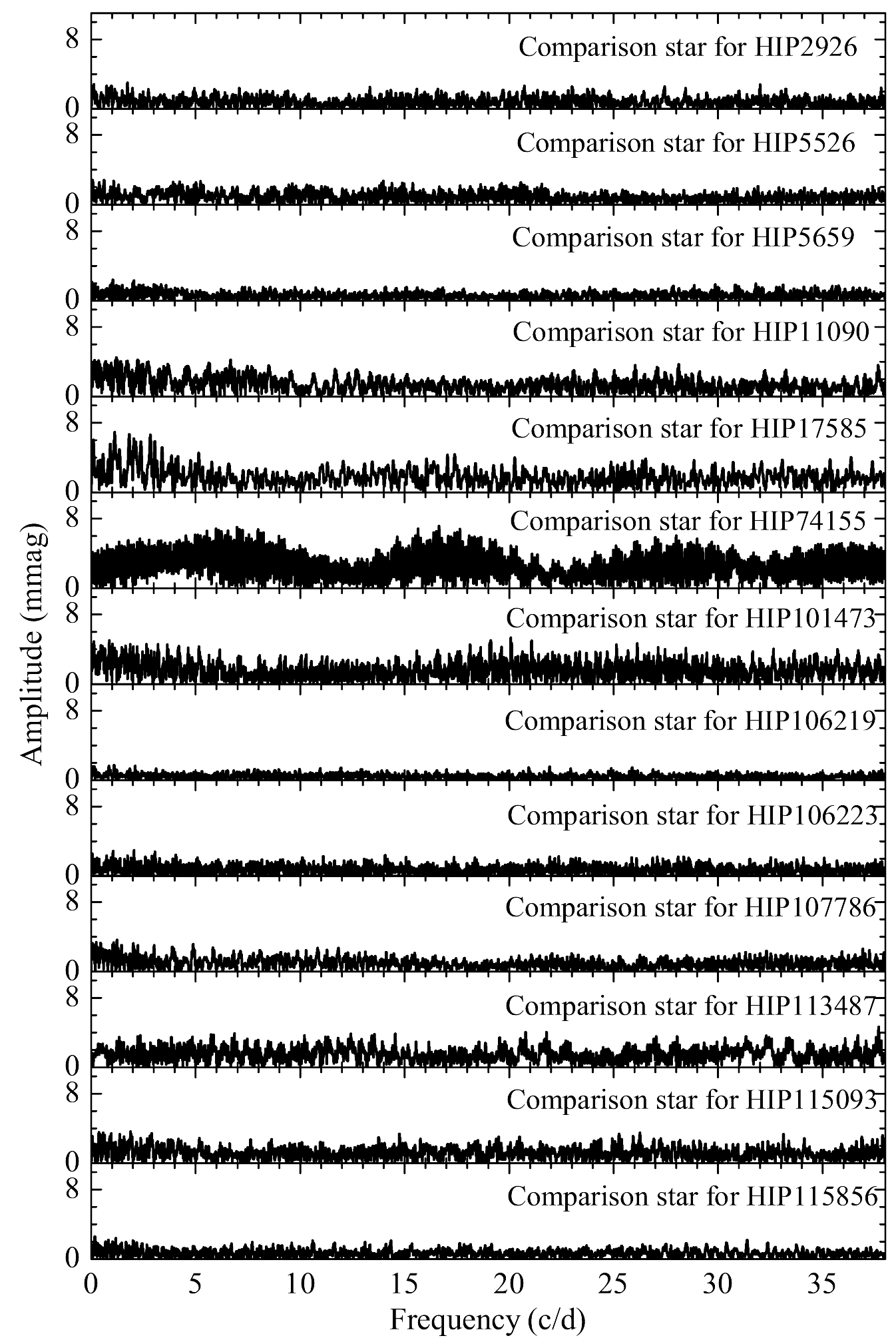}
      \caption{{Amplitude spectra of the comparison stars used for differential photometry of $\delta$\,Scuti candidates. } 
      }  
         \label{Fig3_FT_compar}
   \end{figure}

For the further analysis, we calculated differential magnitudes of our targets using comparison stars. We used 
one comparison star per field, which had a magnitude most similar to the target magnitude and which had a light curve with no signs of variability. The names of used comparison stars are listed in the last column of Table~\ref{table:1}.  We obtained amplitude spectra of selected comparison stars (Figure~\ref{Fig3_FT_compar}) and checked, if there are signals at the same frequencies which were observed in $\delta$\,Scuti candidates.

The LCs were analyzed using a process of their Fourier decomposition into sinusoidal components (\citealt{Fourier1822}).
We used a software Period04\footnote{https://www.univie.ac.at/tops/Period04/}  \citep{Lenz05} for decomposition of LCs, obtaining 
amplitude spectra and for prewhitening procedures in order to find all frequencies, amplitudes and phases of pulsations in light curves, spectral windows (SW) and a noise level. As one site observations were used for analysis, SWs have high side-lobes of 1\,$c/d$ aliases. The highest side-lobes of 1\,$c/d$ aliases were calculated for HIP\,74155 and reach 95.7$\%$ of the central peak, while the lowest one was detected for HIP\,115856 and is equal to 83.1$\%$ of the central peak. The length of the LCs also differs, thus the FWHM of the central lobe in SWs usually vary between 0.0374\,$c/d$ (HIP\,106223) and 0.1558\,$c/d$ (HIP\,11090). The worst SW was obtained for HIP\,74155, since it had a smallest set of data points and big ($2-5$ days) gaps between runs. 

Some stars in their amplitude spectra had a signal at low frequencies ($\approx$\,1\,$c/d$), which needed a special treatment. Signals caused by a daily variation of weather conditions may appear at such frequencies, especially when a target and a comparison star are of different colors, because of instrumental instabilities or a varying position of a star on a CCD chip during blind tracking, if a star periodically crosses the same defected pixel or dust spot. In cases of significant signals at low frequencies we checked the resulting amplitude spectra using different comparison stars of different colors. 
If analyzis with all comparison stars gave the same signal at low frequencies, we attributed it  to stellar pulsations.    

 We observed the stars in a mode of blind tracking, that might cause artifacts in amplitude spectra at any frequency. Though positions of the stars were not stable from one night to another and during the same night, these variations were not periodic and could not produce periodic signals. Moreover, we performed a careful reduction of the CCD images using calibration images (bias, dark and flat-field) taken for every night separately trying to eliminate any newly appeared dust grain or other defects in the field.

We used a so-called prewhitening procedure for analysis of every star. First of all we calculated an amplitude spectrum with Period04 and identified the highest amplitude peak at frequencies higher than 2\,$c/d$ assuming that the low frequencies may be caused by instrumental or weather instabilities. 
If there were such signals, they were analyzed the last. An exception was done only for two stars: HIP\,2923, as the signal at 0.9651\,$c/d$ was dominant and its side-lobes could affect other signals in the amplitude spectrum; and HIP\,106223, as this star showed signals only at low frequencies.  
After that we calculated a sinusoid with the identified frequency and used a least square fitting method improving amplitude and phase, simultaneously.
Then we checked a significance of the extracted frequency by comparing an amplitude of the signal with the mean amplitude of residual in a box $\pm$10\,$c/d$ around the extracted frequency, i.e. we calculated a signal to noise ratio (S/N) at the extracted frequency. The noise level was calculated using the same software Period04. According to \cite{Breger1993}, a signal may be assumed as significant if its ${\rm S/N} \gtrapprox 4$. \citet{Alvarez1998} have shown that signals with ${\rm S/N}=3.7$ in a box of 10~$c/d$ could be as an indication of 99$\%$ of significance level, while ${\rm S/N}=3.2$ is an indication of 90$\%$ of significance level. 
Some authors use a S/N cut as high as 6 in order to be 100$\%$ confident in significance of signals.  Almost a half of our analyzed stars have at least one frequency reaching the ${\rm S/N} \geq 6$ level. Another five analyzed stars did not exceed ${\rm S/N} = 5$ level. We interpreted signals as significant when their ${\rm S/N} \geq 4$ in a box of $\pm$10~$c/d$ around the extracted frequency, and verified if changes in the box size by up/down 10\,$c/d$ do not
alter the S/N level significantly. We listed the extracted frequencies according to their extraction succession, thereby experienced readers may make their own decision whether to accept a given frequency or not.
We extracted only one signal at frequencies lower than 2~$c/d$ even if peaks left after the extraction gave ${\rm S/N} >4$, because it is always difficult to deal with so low frequencies from ground-based observations. In that case we just state the fact of low frequency detection without its detailed analyzis.

\begin{table}
\caption{Observed signals in amplitude spectra.} 
\label{table:2}      
\resizebox{\columnwidth}{!}{
\centering                          
\begin{tabular}{c c c c c}        
\hline
             Frequency $\pm \sigma$      &      Amplitude $\pm \sigma$	& Phase $\pm \sigma$		   &  Noise & 	S/N    \\ 
            $c/d$	& mmag	&      &  mmag  &	       \\
\hline
\multicolumn{5}{c}{HIP\,2923}  \\
   0.9651$\pm$0.0022  &   8.73$\pm$0.62 &  0.227$\pm$0.015 &  1.66&	5.25   \\
          15.0271$\pm$0.0019 &   8.41$\pm$0.65 &  0.764$\pm$0.010 &  1.45&	5.81	  \\
          16.0973$\pm$0.0022 &   7.28$\pm$0.66 &  0.277$\pm$0.012 &  1.25&	5.84	  \\
          11.7260$\pm$0.0021 &   6.24$\pm$0.62 &  0.987$\pm$0.013 &  1.26&	4.94	  \\
          6.7078$\pm$0.0025  &   6.02$\pm$0.61 &  0.845$\pm$0.015 &  1.40&	4.30   \\
          11.4043$\pm$0.0032 &   4.71$\pm$0.61 &  0.262$\pm$0.021 &  1.09 &	4.32   \\
\multicolumn{5}{c}{HIP\,5526}          \\
   9.3431$\pm$0.0007  &  24.19$\pm$0.77 & 0.806$\pm$0.005  &   4.77&	5.07	\\	
          5.1385$\pm$0.0009  &  20.46$\pm$0.77 & 0.557$\pm$0.007  &   4.23&	4.84	  \\
          8.5147$\pm$0.0007  &  17.92$\pm$0.83 & 0.788$\pm$0.005  &   3.26&	5.50	\\	
          9.7283$\pm$0.0012  &  13.21$\pm$0.78 & 0.291$\pm$0.008  &   2.70&	4.89	\\	
          5.5947$\pm$0.0014  &  12.46$\pm$0.84 & 0.849$\pm$0.010  &   2.58&	4.82	  \\
          12.5515$\pm$0.0021 &  8.22$\pm$0.80  & 0.255$\pm$0.014  &   1.86&	4.41	  \\
          0.542$\pm$0.0019   &  8.05$\pm$0.76  & 0.799$\pm$0.013  &   1.75&	4.61	\\	
\multicolumn{5}{c}{HIP\,5659}              \\
   9.4932$\pm$0.0164  &  16.45$\pm$0.94 &   0.049$\pm$0.011&   2.90&	5.67   \\
          10.2439$\pm$0.0115 &  11.49$\pm$0.85 &   0.629$\pm$0.014&   2.45&	4.69   \\
          9.8508$\pm$0.0125  &  12.46$\pm$0.93 &   0.674$\pm$0.012&   1.98&	6.30   \\
          7.0028$\pm$0.0089  &   8.95$\pm$0.83 &   0.814$\pm$0.015&   1.82&	4.92   \\
\multicolumn{5}{c}{HIP\,11090}                   \\
  15.8617$\pm$0.0024 &  11.30$\pm$1.13 &  0.387$\pm$0.016 &  1.63&	6.94   \\
          28.7418$\pm$0.0049 &  5.34$\pm$.59  &  0.453$\pm$0.034 &  1.19&	4.47   \\
           0.9662$\pm$0.0027 &  8.89$\pm$.80  &  0.811$\pm$0.019 &  1.93&	4.61   \\
\multicolumn{5}{c}{HIP\,17585}               \\
  13.1631$\pm$0.0078 &  13.99$\pm$2.01 &  0.887$\pm$0.023 &  3.33&	4.20   \\
 \multicolumn{5}{c}{HIP\,74155}                  \\
  11.7619$\pm$0.0039 &  14.15$\pm$2.31 &  0.847$\pm$0.026 &  2.99&	4.73   \\
\multicolumn{5}{c}{HIP\,101473}                 \\
 6.0374$\pm$0.0063 &   7.24$\pm$1.44 &  0.997$\pm$0.033 &  1.79&	4.04   \\
 4.3081$\pm$0.0080 &   5.64$\pm$1.57 &  0.038$\pm$0.042 &  1.34& 4.20   \\
 \multicolumn{5}{c}{HIP\,106219}                  \\
 11.3007$\pm$0.0011 &   6.66$\pm$0.44 &  0.113$\pm$0.010 &  0.74&	8.95	  \\
          10.8018$\pm$0.0023 &   2.91$\pm$0.43 &  0.042$\pm$0.023 &  0.69&	4.19	  \\
          14.2773$\pm$0.0027 &   2.45$\pm$0.42 &  0.210$\pm$0.027 &  0.52  & 4.74	    \\
\multicolumn{5}{c}{HIP\,106223}                 \\
 1.1429$\pm$0.0004  &  31.09$\pm$0.71 &  0.194$\pm$0.004 &  2.84&	10.95  \\
 \multicolumn{5}{c}{HIP\,107786}                  \\
 15.4817$\pm$0.0025 &   9.87$\pm$1.53 &  0.932$\pm$0.025 &  2.08&	4.75   \\
          1.3778$\pm$0.0018  &  13.11$\pm$1.48 &  0.476$\pm$0.018 &  2.78  & 4.72     \\
\multicolumn{5}{c}{HIP\,113487}       \\
 21.9102$\pm$0.0013 &  23.01$\pm$1.80 &  0.327$\pm$0.013 &  3.73&	6.16	  \\
 17.2064$\pm$0.0025 &  11.40$\pm$1.89 &  0.012$\pm$0.024 &  2.82&	4.04	  \\
\multicolumn{5}{c}{HIP\,115093}  \\
 11.4318$\pm$0.0040 &  10.61$\pm$1.39 &  0.784$\pm$0.021 &  2.18&	4.87	  \\
\multicolumn{5}{c}{HIP\,115856}        \\
 9.1109$\pm$0.0007  &  14.08$\pm$0.65 &  0.917$\pm$0.008 &  1.75&	8.06	\\	
          16.9660$\pm$0.0016 &   7.48$\pm$.69 &  0.068$\pm$.016 &  1.28&	5.86	  \\
          18.6810$\pm$0.0018 &   5.81$\pm$.68 &  0.676$\pm$.019 &  1.11&	5.24	  \\
          0.9669$\pm$0.0011  &   9.55$\pm$.00 &  0.240$\pm$.011 &  1.29&	7.43   \\    
\hline                                   
\end{tabular}
}
\end{table}

\begin{figure*}
   \centering
   \includegraphics[width=15cm]{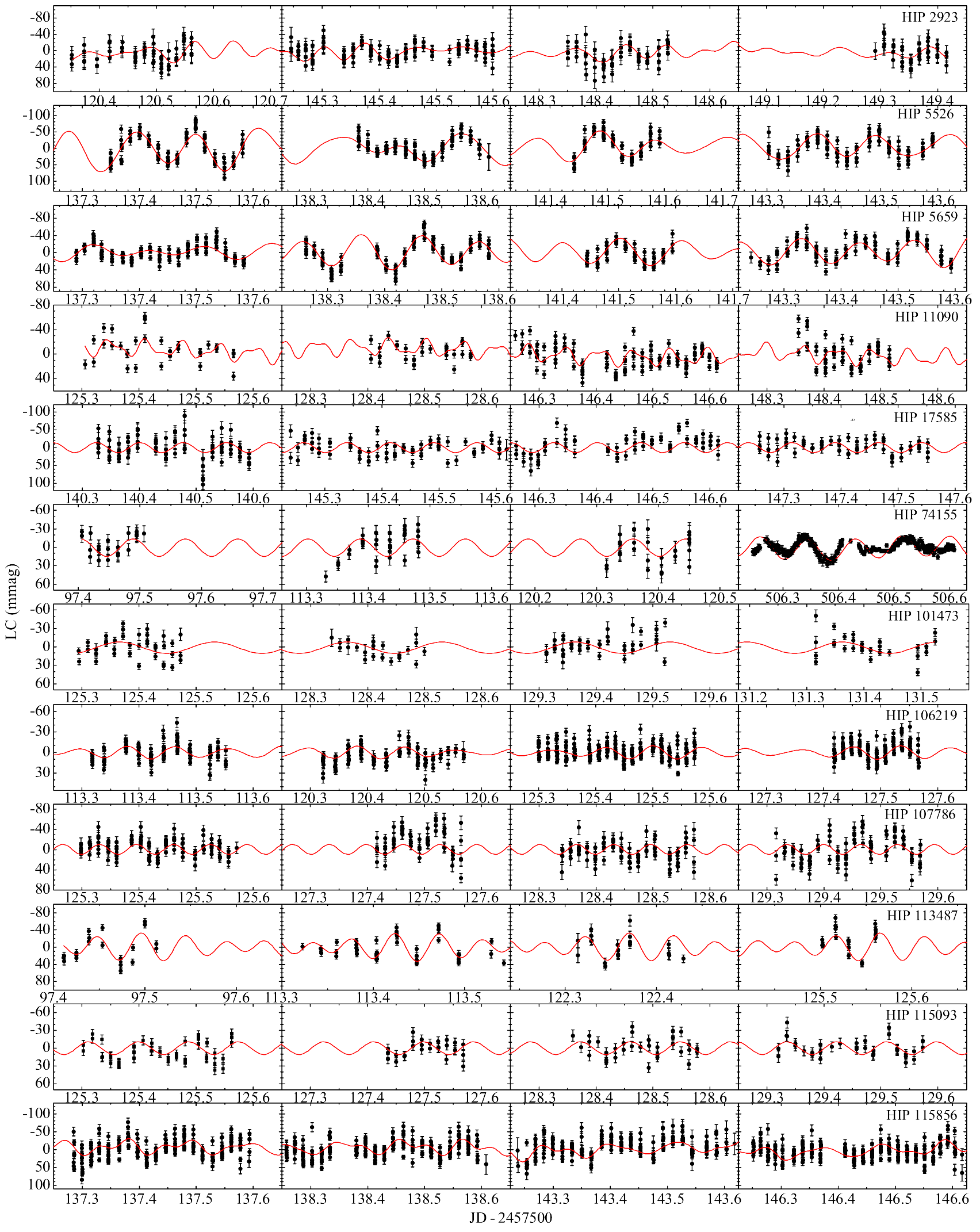}
   \caption{Light curves of the $\delta$\,Scuti candidate stars obtained during four best quality observing runs. The black dots with error-bars correspond to observations, the red curves represent the LC calculated according to the determined frequencies and amplitudes of pulsations. 
   }
   \label{Fig4_LC_fragment}
    \end{figure*}  
    
The upper bound of frequency is defined by the Nequest frequency, which depends on the sampling time of light curves. In our case with an uneven sampling time, the effective Nequest frequency is close to 50\,$c/d$, which is twice larger than the highest frequency of pulsations observed in our sample of stars.

\section{Characterization of $\delta$\,Scuti candidates}

We have collected enough data for the variability analyzis of the 13 $\delta$\,Scuti candidates  (Table~\ref{table:1}).  
As it was expected, the amplitude spectra of LCs revealed signals mostly between 5\,$c/d$ and 22\,$c/d$. This range of frequencies is intrinsic for $\delta$\,Scuti stars. Some of those stars had pulsations at frequencies below 2\,$c/d$ and this may be an indication of $\gamma$\,Doradus or $\delta$\,Scuti-$\gamma$\,Doradus hybrid star. However, we have to consider that the frequency at around 1\,$c/d$ may be caused by the daily atmospheric transmittance variations or instrumental instabilities.

A list of all detected signals ordered according to their prewhitening in every star is presented in Table~\ref{table:2}.  
Figure~\ref{Fig4_LC_fragment} shows how theoretical and observed LCs fit each other. The observed LCs correspond to four best quality runs for every star. The theoretical LCs were calculated using the computing program Period04 \citep{Lenz05} with the parameters of brightness variations given in Table~\ref{table:2}.

Below we discuss every star separately in more detail.

\subsection{HIP\,2923}

    \begin{figure}[!ht]
   \centering
   \includegraphics[width=\hsize]{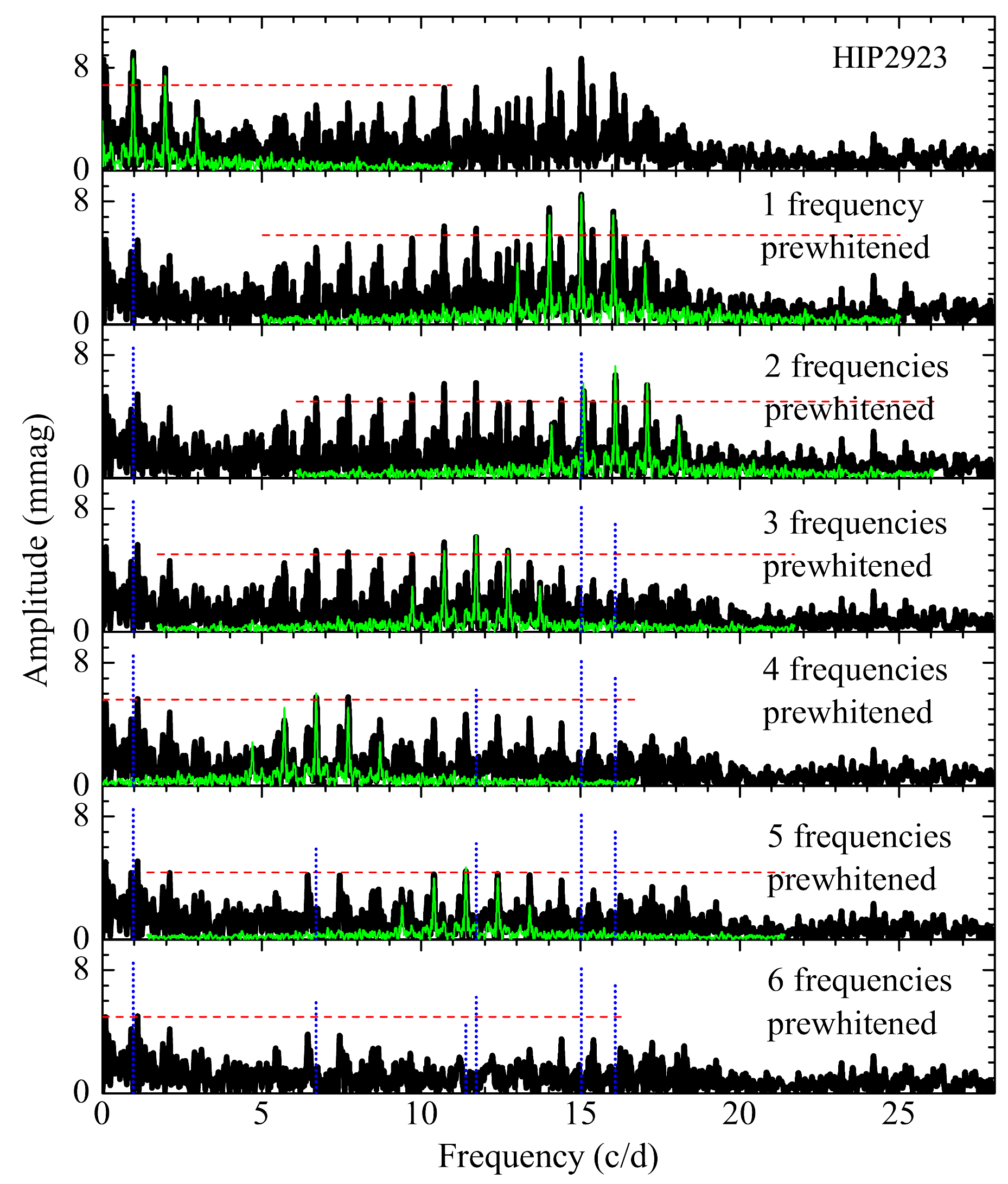}
      \caption{Amplitude spectrum of the combined LC of HIP\,2923 and its prewhitening. In each panel, from top to bottom, one signal with S/N$>$4 is prewhitened from the time series, and a new spectrum of residuum is obtained.     
      The black solid curves show amplitude spectra. The selected for prewhitening signal has a shape of the spectral window, which is drawn on top of the amplitude spectrum.
      The dashed horizontal line corresponds to a 4 times of noise level at the selected frequency. Length of the dashed horizontal line shows size of the box used for the noise level calculation. The vertical dotted lines show positions and amplitudes of already prewhitened frequencies.
      }  
         \label{Fig5_FT_2923}
   \end{figure}
   
 According to fundamental parameters derived by \cite{McDonald2012}, HIP\,2923 is located on a blue edge of $\gamma$\,Doradus stars defined by \cite{Dupret2005} (Figure~\ref{Fig1_strip}).  An automated classification of Hipparcos unsolved variables by \cite{Rimoldini2012} classified it as a low amplitude $\delta$\,Scuti variable: both {\it Random forest} (RF) and {\it Multistage Bayesian Network} (MB) methods gave the same result with probabilities of 0.86 and 0.89,  respectively. \cite{Rimoldini2012} also derived a frequency of the dominant signal to be 9.3455~$c/d$.
 
 The prewhitening process of our data on HIP\,2923 is presented in Figure~\ref{Fig5_FT_2923}. 
 An amplitude spectrum of HIP\,2923 is rich in signals.  
 We found two dominant signals at 0.9651\,$c/d$ and 15.0271\,$c/d$ with amplitudes 8.73\,mmag and 8.41\,mmag, respectively. That may be an indication of the $\delta$\,Scuti--$\gamma$\,Doradus hybrid star pulsation. In total, we found five signals at frequencies common to $\delta$\,Scuti type stars and one signal at low frequencies typical to $\gamma$\,Doradus type stars.
 As the signal at 0.9651\,$c/d$ needed a special caution, we checked the resulting amplitude spectra using different comparison stars of different colors, but no one of them reduced its amplitude. 
 
Taking into account the position of HIP\,2923 in the HR diagram and detected signals of pulsations HIP\,2923 could be considered as a $\delta$\,Scuti--$\gamma$\,Doradus hybrid star.
 
\subsection{HIP\,5526}

   \begin{figure}[!ht]
   \centering
   \includegraphics[width=\hsize]{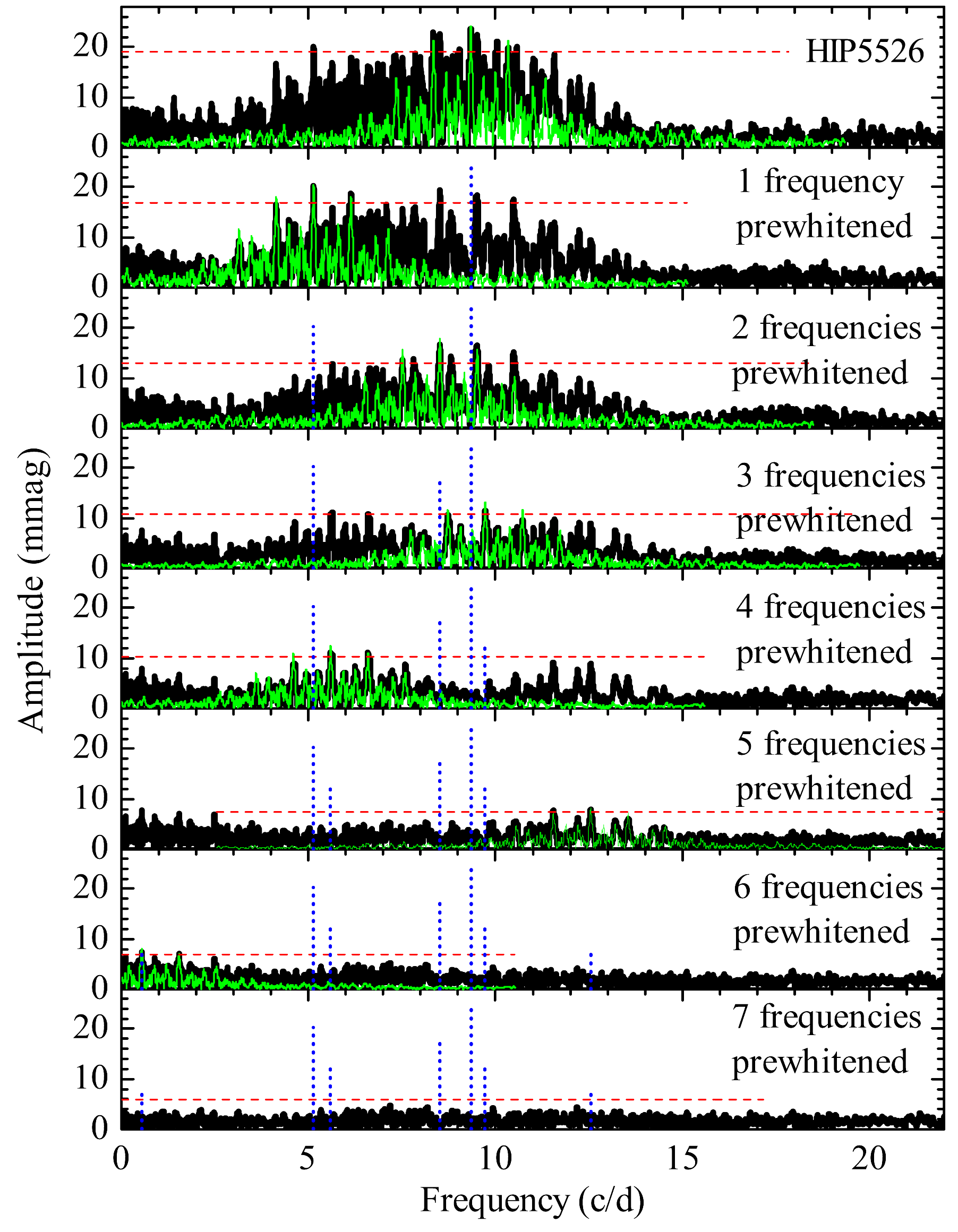}
      \caption{Amplitude spectrum of the combined LC of HIP\,5526, and its prewhitening. Meanings of the lines are the same as in Figure~\ref{Fig5_FT_2923}.
      }  
         \label{Fig6_FT_5526}
   \end{figure}

According to fundamental parameters derived by \cite{McDonald2012}, HIP\,5526 is close to the red edge of $\gamma$\,Doradus instability strip derived by \cite{Dupret2005}.  An automated classification of Hipparcos unsolved variables by \cite{Rimoldini2012} classified HIP\,5526 as a low amplitude $\delta$\,Scuti variable with probabilities equal to 0.28 (RF method) and 0.46 (MB method). 

\cite{Rimoldini2012} used Hipparcos observations and found for this star a dominant signal at 0.67842\,$c/d$. We find more frequencies of pulsations for this star in our set of observations. 
We prewhitened seven frequencies from LC of HIP\,5526 (Table~\ref{table:2} and Figure~\ref{Fig6_FT_5526}). Those signals could not come from the used comparison stars, as amplitude spectra of comparison stars do not show signs of variability (Figure~\ref{Fig3_FT_compar}). The dominant signal in the amplitude spectrum of HIP\,5526 is at 9.3431\,$c/d$. As well as \cite{Rimoldini2012}, we found that HIP\,5526 has a signal at low frequency. It peaks at 0.5420\,$c/d$ with $\rm{S/N}=4.61$. 
According to position in the HR diagram and the signal at low frequency, HIP\,5526 could be  considered as a $\delta$\,Scuti--$\gamma$\,Doradus hybrid star. This presumption should be taken into acount in further analyses of HIP\,5526.

\subsection{HIP\,5659}

   \begin{figure}[!h]
   \centering
   \includegraphics[width=\hsize]{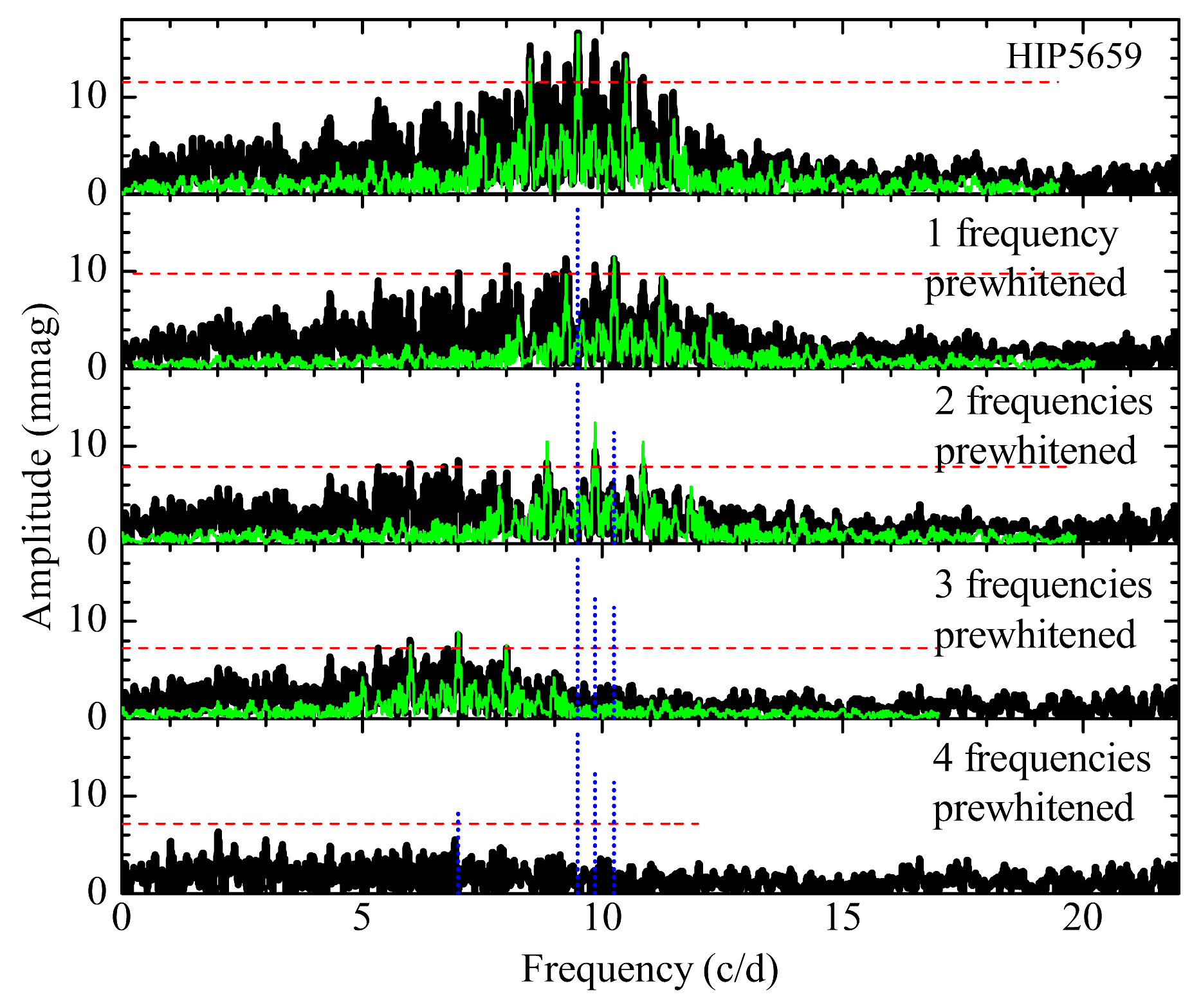}
      \caption{Amplitude spectrum of the combined LC of HIP\,5659, and its prewhitening. Meanings of the lines are the same as in Figure~\ref{Fig5_FT_2923}.
      }  
         \label{Fig7_FT_5659}
   \end{figure}

According to fundamental parameters derived by \cite{McDonald2012}, HIP\,5659 is close to the red edge of $\gamma$\,Doradus instability strip derived by \cite{Dupret2005}. In Simbad database, HIP\,5659 is classified as a double or multiple star.
\cite{Rimoldini2012} classified HIP\,5659 as low amplitude $\delta$\,Scuti variable with probabilities equal 0.57 (RF method) and 0.69 (MB method). 

 We prewhitened 4 frequencies from LC of HIP\,5659 (Table~\ref{table:2}, Figure~\ref{Fig7_FT_5659}). We found the dominant signal at 9.4932\,$c/d$ which is very close to the dominant signal 9.85145\,$c/d$ found by \cite{Rimoldini2012}, who used the Hipparcos data. The amplitude spectrum of HIP\,5659 looks typical to $\delta$\,Scuti type star and does not show any sign of variability at frequencies below 2\,$c/d$.

\subsection{HIP\,11090}

   \begin{figure}[!h]
   \centering
   \includegraphics[width=\hsize]{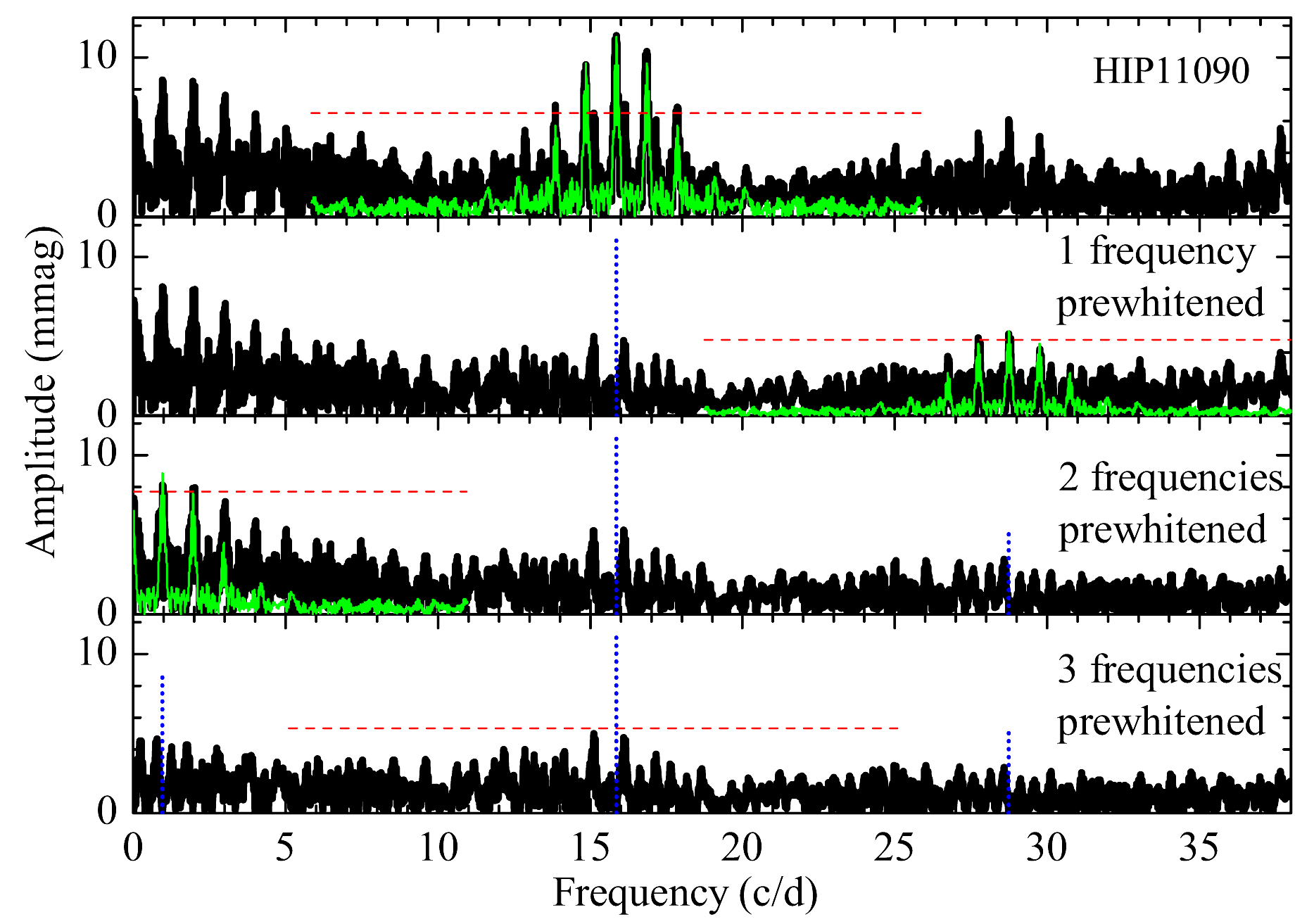}
      \caption{Amplitude spectrum of the combined LC of HIP\,11090 and its prewhitening. Meanings of the lines are the same as in Figure~\ref{Fig5_FT_2923}. 
      }  
         \label{Fig8_FT_11090}
   \end{figure}

According to fundamental parameters derived by \cite{McDonald2012}, HIP\,11090 is located in the center of the instability strip of $\gamma$\,Doradus stars defined by \cite{Dupret2005} (Figure~\ref{Fig1_strip}).  \cite{Rimoldini2012} classified this star as a low amplitude $\delta$\,Scuti variable with probabilities of 0.34 (RF method) and 0.52 (MB method). \cite{Rimoldini2012} also derived the dominant signal at 15.86354\,$c/d$.

We observed HIP\,11090 during eleven runs and obtained a really nice amplitude spectrum with an obvious peak at 15.8617\,$c/d$ with amplitude of 11.30~mmag (Figure~\ref{Fig8_FT_11090}, Table~\ref{table:2}). This frequency is really close to the one derived by \cite{Rimoldini2012}. In addition, we found two more signals with ${\rm S/N}>4$ at 28.7418\,$c/d$ and 0.9662\,$c/d$. We tend to trust the low frequency signal in HIP\,11090 due to its position in the HR diagram suggesting that this star may be a $\gamma$\,Doradus star or a hybrid star. 
As HIP\,11090 shows pulsations at frequencies typical to both $\delta$\,Scuti and $\gamma$\,Doradus, this star could be considered as a hybrid star.

\subsection{HIP\,17585}

   \begin{figure}[!h]
   \centering
   \includegraphics[width=\hsize]{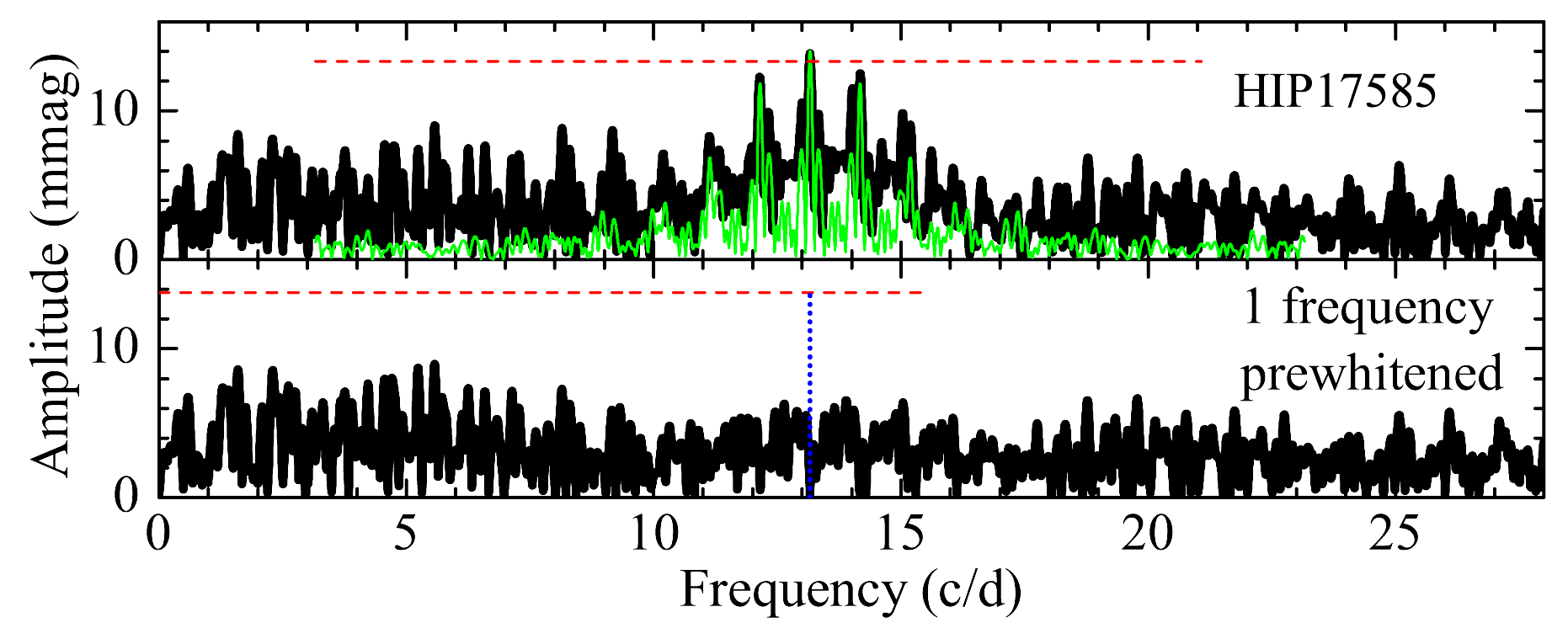}
      \caption{Amplitude spectrum of the combined LC of HIP\,17585 and its prewhitening. Meanings of the lines are the same as in Figure~\ref{Fig5_FT_2923}.  
      }  
         \label{Fig9_FT_17585}
   \end{figure}

According to fundamental parameters derived by \cite{McDonald2012}, HIP\,17585 is located at the red edge of the instability strip of $\gamma$\,Doradus stars defined by \cite{Dupret2005} (Figure~\ref{Fig1_strip}).
\cite{Kahraman2016} used FIES spectra obtained on the Nordic Optical Telescope for spectroscopic analysis of atmospheric parameters and abundance of $\gamma$\,Doradus stars, and classified HIP\,17585 (another name is HD\,23005) as a $\gamma$\,Doradus candidate of F1\,IV spectral type. 
\cite{Rimoldini2012} found pulsations at frequency 1.64172\,$c/d$ and classified HIP\,17585 as a low amplitude $\delta$\,Scuti variable with probability of 0.46 using the RF method, or as a $\gamma$\,Doradus star with probability of 0.42 using the MB method. 

The amplitude spectrum of LC observed by us showed only one peak with ${\rm S/N}>4$ at 13.1631\,$c/d$ with amplitude of 13.99\,mmag (Figure~\ref{Fig9_FT_17585} and Table~\ref{table:2}). HIP\,17585 has a higher noise level in its amplitude spectrum  than other stars discussed before. Probably this is a reason why we could not detect any signal  common in $\gamma$\,Doradus stars. Even the noise at low frequencies observed in the amplitude spectrum of the comparison star (Figure~\ref{Fig3_FT_compar}) did not show up in the amplitude spectrum of HIP\,17585. Anyway, we can conclude that HIP\,17585 has pulsations at frequencies common in $\delta$\,Scuti stars.

\subsection{HIP\,74155}

   \begin{figure}[!h]
   \centering
   \includegraphics[width=\hsize]{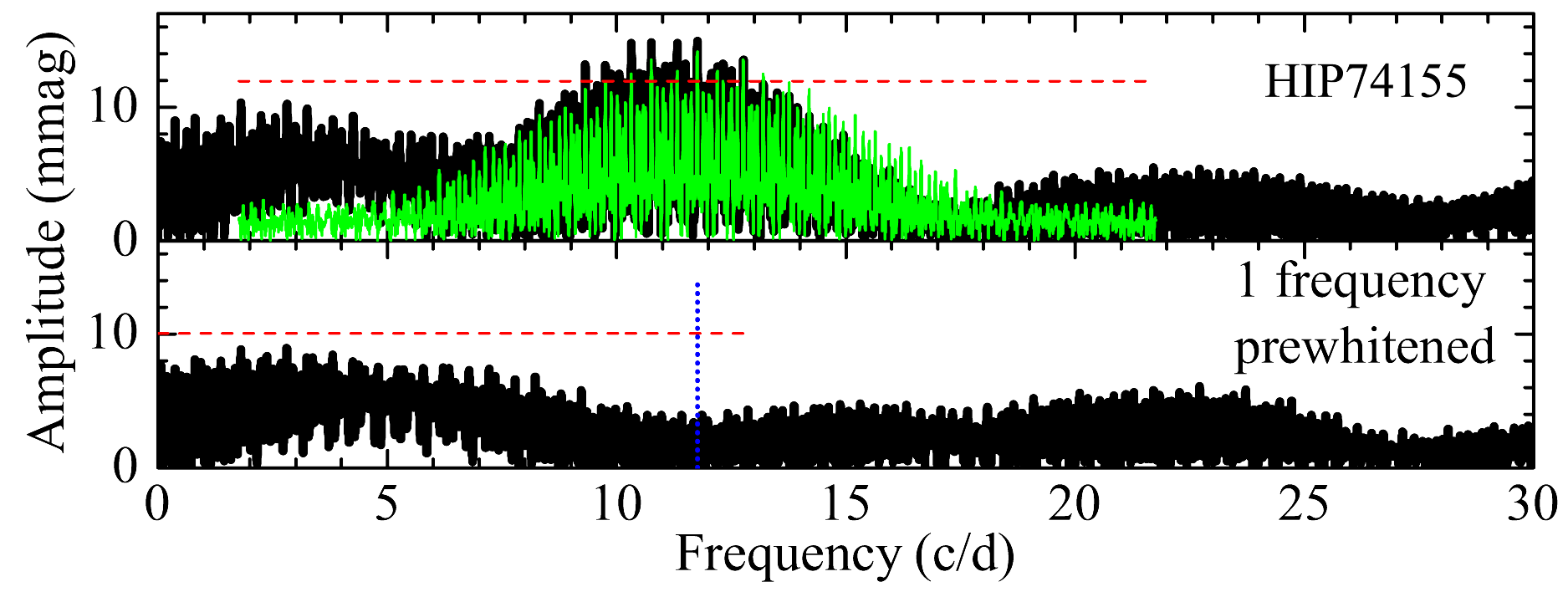}
      \caption{Amplitude spectrum of the combined LC of HIP\,74155 and its prewhitening. Meanings of the lines are the same as in Figure~\ref{Fig5_FT_2923}. }  
         \label{Fig10_FT_74155}
   \end{figure}

According to fundamental parameters derived by \cite{McDonald2012}, HIP\,74155 is located outside 
of the $\gamma$\,Doradus instability strip defined by \cite{Dupret2005}, however still inside the $\gamma$\,Doradus instability strip defined by \cite{Xiong2016} (Figure~\ref{Fig1_strip}).
\cite{Rimoldini2012} classified HIP\,74155 as a low amplitude $\delta$\,Scuti variable with probabilities of 0.55 (RF method) and 0.60 (MB method). \cite{Rimoldini2012} also derived frequency of the dominant signal, which was 10.20234\,$c/d$.
              
Our observations of HIP\,74155 had quite big gaps of different length ($2-5$~days). This  disturbed the spectral window and raised difficulties in prewhitening of signals. We detected only one signal at 11.7619\,$c/d$ with amplitude of 14.15~mmag and ${\rm S/N}>4$ (Figure~\ref{Fig10_FT_74155} and Table~\ref{table:2}). 

In order to collect more data on variability of this star, we observed HIP\,74155 not just during five runs on the Maksutov-type telescope in 2016, but also involved our larger 1.65~m MAO telescope in 2017. The additional observations showed clear variability in the range of $\delta$\,Scuti pulsation frequencies. Also we found that HIP\,74155 has more than one frequency of pulsations (see Figure~\ref{Fig4_LC_fragment} between JD\,2458006.23 and JD\,2458006.62). As we observed only one run with the 1.65~m MAO telescope, S/N of signals in the amplitude spectrum was smaller than 3.3 (i.e. below the limit we set for our detailed analysis), thus the determined two signals at 11.059~$c/d$ with an amplitude of 9.4~mmag and at 14.515~$c/d$ with the amplitude of 6.7~mmag we present here just as  indications for further analyses. 
 However, our observations allow us to conclude that HIP\,74155 has more than one frequency of pulsations which are typical to a $\delta$\,Scuti type variable stars.

\subsection{HIP\,101473}

   \begin{figure}[!h]
   \centering
   \includegraphics[width=\hsize]{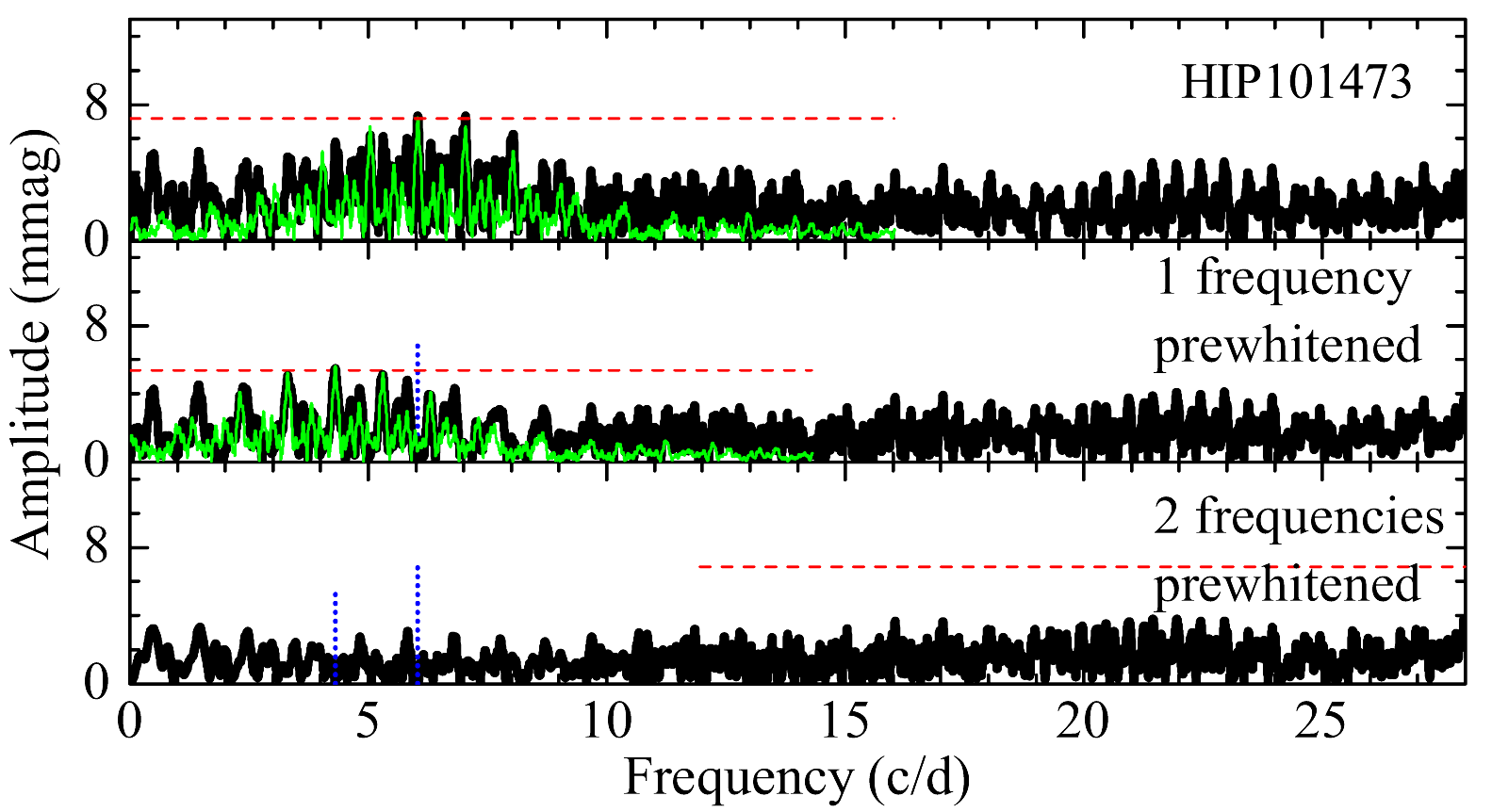}
      \caption{Amplitude spectrum of the combined LC of HIP\,101473 and its prewhitening. Meanings of the lines are the same as in Figure~\ref{Fig5_FT_2923}.  }  
         \label{Fig11_FT_101473}
   \end{figure}

HIP\,101473 is the hottest star in our sample. According to fundamental parameters derived by \cite{McDonald2012}, it appears at the blue edge of instability strip of $\delta$\,Scuti stars derived by \cite{Dupret2005} and \cite{Xiong2016} (Figure~\ref{Fig1_strip}). 
\cite{Rimoldini2012} classified HIP\,101473 as a low amplitude $\delta$\,Scuti variable with probabilities of 0.39 (RF method) and 0.43 (MB method). \cite{Rimoldini2012} also derived  frequency of the dominant signal, which was 7.50067\,$c/d$.

This star was observed by us during seven runs. Some of the runs showed a wavy shape of LCs inherent to variable stars (Figure~\ref{Fig4_LC_fragment}), however there were runs without obvious brightness variations. Analysis of combined LC of all seven runs gave two signals at frequencies of 6.0374\,$c/d$ and 4.3081\,$c/d$ with S/N just slightly larger than 4 (Figure~\ref{Fig11_FT_101473} and Table~\ref{table:2}). 
The value of the higher frequency is close to the one derived by \cite{Rimoldini2012}. 
We confirm that HIP\,101473 has pulsations at frequencies typical to $\delta$\,Scuti type stars, however it needs further observations in order to determine its pulsation parameters more precisely.

\subsection{HIP\,106219}

   \begin{figure}[!h]
   \centering
   \includegraphics[width=\hsize]{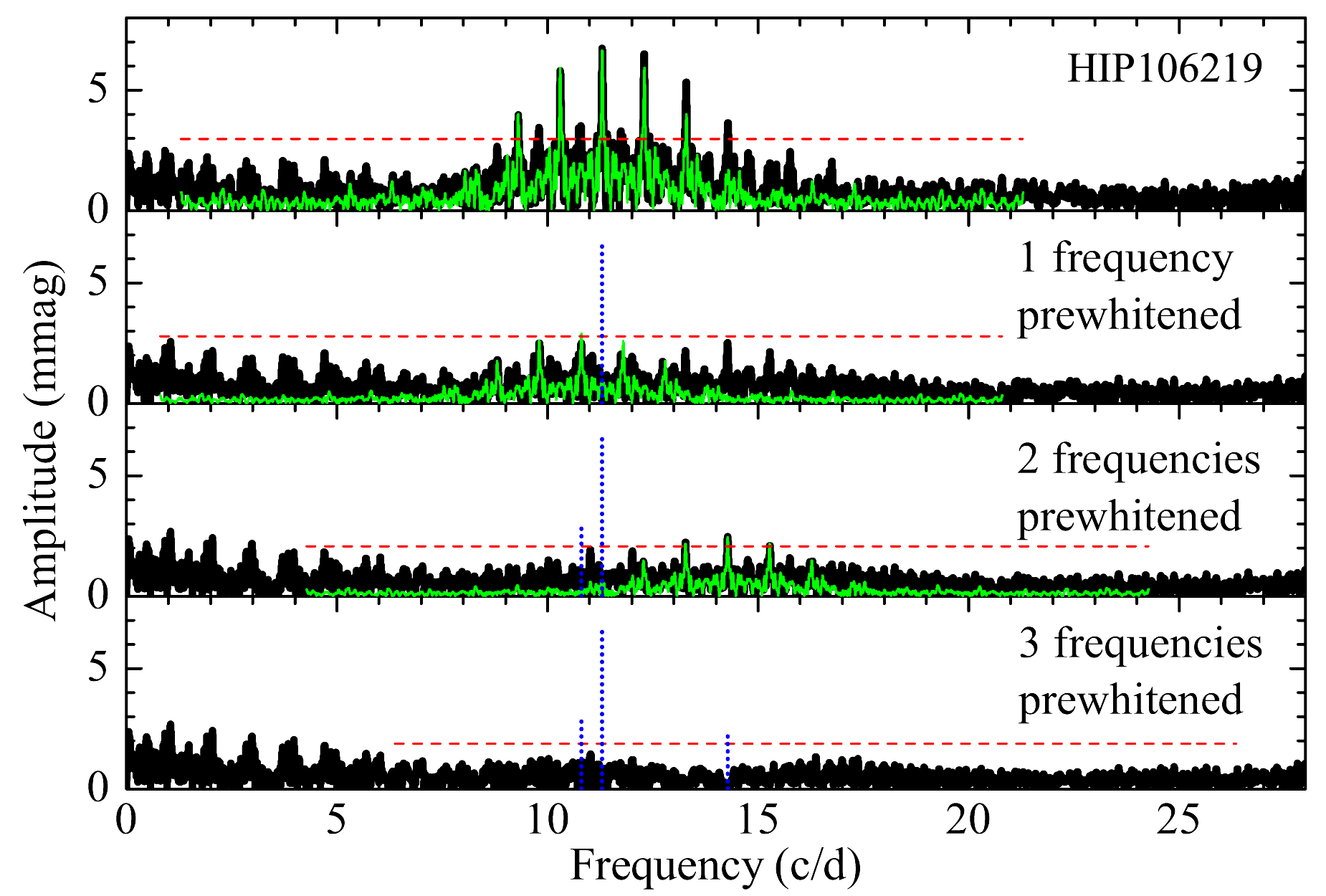}
      \caption{Amplitude spectrum of the combined LC of HIP\,106219 and its prewhitening. Meanings of the lines are the same as in Figure~\ref{Fig5_FT_2923}.  }  
         \label{Fig12_FT_106219}
   \end{figure}

According to fundamental parameters derived by \cite{McDonald2012}, HIP\,106219 appiers in the HD diagram in the middle of $\delta$\,Scuti instability strip, which overlaps with instability strip of $\gamma$\,Doradus stars (\citealt{Xiong2016}). This star is one of the most evolved stars among our targets (Figure~\ref{Fig1_strip}). 
\cite{Rimoldini2012} classified HIP\,106219 as a low amplitude $\delta$\,Scuti variable with probabilities of 0.53 (RF method) and 0.88 (MB method). \cite{Rimoldini2012} also derived  frequency of the dominant signal, which was 10.80438\,$c/d$.

We have observed HIP\,106219 during 14 runs, thus a noise level in amplitude spectrum of this star was smallest among 13 observed targets due to very good weather conditions.  The light curve of HIP\,106219 observations gave a very nice amplitude spectrum with the obvious signal at 11.3007\,$c/d$ with amplitude of 6.66~mmag and two lower amplitude signals at 10.8018\,$c/d$ and 14.2773\,$c/d$  (Figure~\ref{Fig12_FT_106219} and Table~\ref{table:2}). One of our frequencies fits quite well with the frequency derived by \cite{Rimoldini2012}.

The amplitude spectrum of HIP\,106219 also has an enlarged amplitude at low frequencies with the highest peak at 1.0468\,$c/d$ with amplitude of 2.70\,mmag  and with ${\rm S/N}=4.17$. As the instability strip of $\gamma$\,Doradus stars is quite wide (Figure~\ref{Fig1_strip}), there is a possibility that we managed to observe low amplitude pulsations at low frequency, which are typical to $\gamma$\,Doradus stars. Presently we do not accept this signal as significant enough, however it should be considered during further analysis of HIP\,106219.   
We conclude that HIP\,106219 has pulsations at frequencies typical to a $\delta$\,Scuti star, however further analyses may reveal it as being a hybrid star.

\subsection{HIP\,106223}

\begin{figure}[!h]
\centering
\includegraphics[width=\hsize]{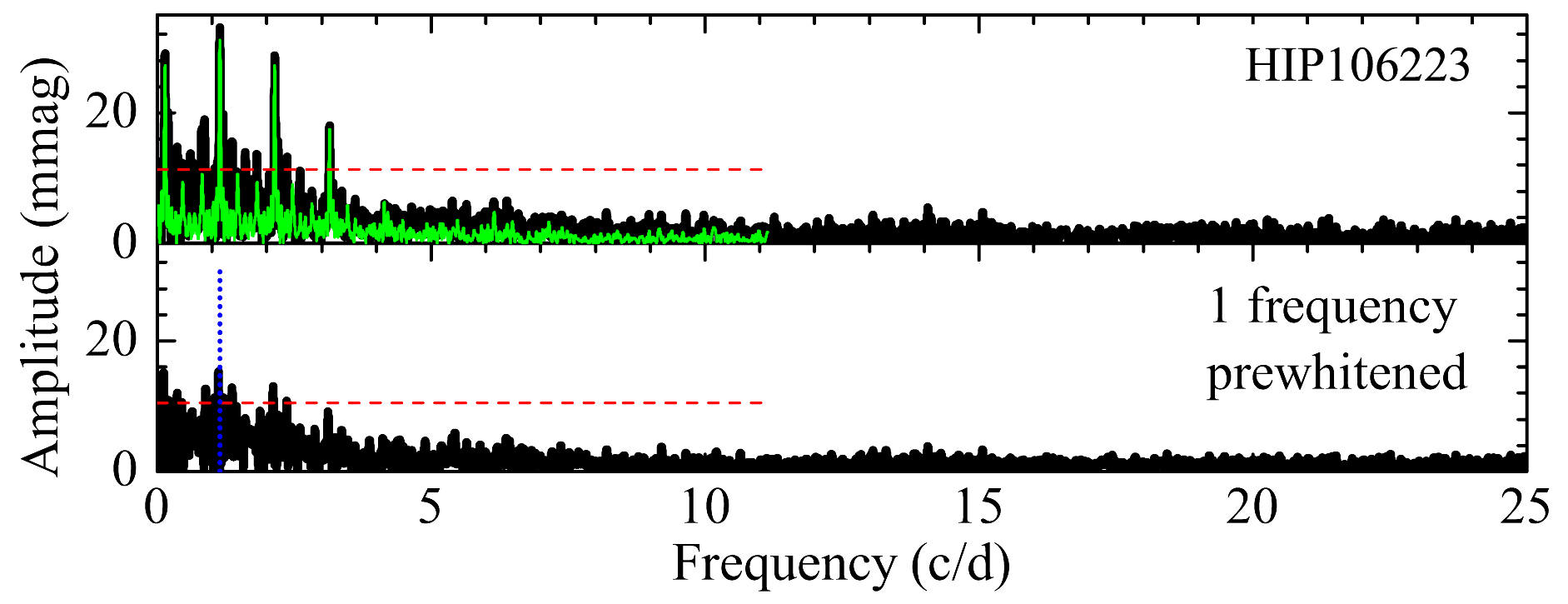}
\caption{Amplitude spectrum of the combined LC of HIP\,106223 and prewhitening process. Meanings of the lines are the same as in Figure~\ref{Fig5_FT_2923}.
}
\label{Fig13_FT_106223}
\end{figure}

\begin{figure}[!h]
   \centering
   \includegraphics[width=\hsize]{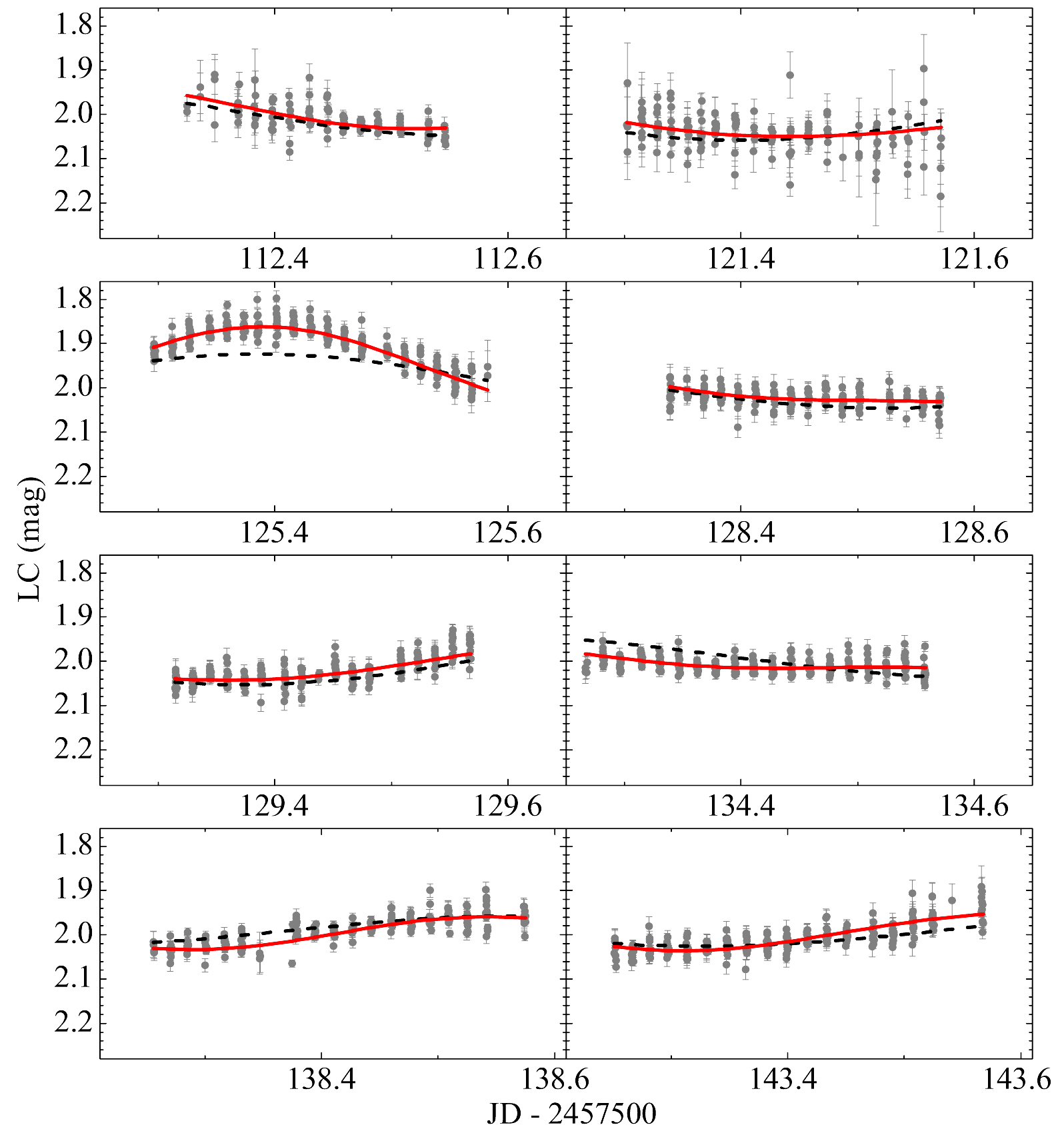}
   \caption{Light curves of HIP\,106223 during eight observing runs. The gray dots with error bars show the observed LC, the continuous and dashed curves show the calculated LCs using 10 and 3 frequencies of pulsations, respectively. 
   }
   \label{Fig14_LC_106223}
    \end{figure}  
    
    According to fundamental parameters derived by \cite{McDonald2012}, HIP\,106223 is located close to the ZAMS at the red edge of the instability strip of $\gamma$\,Doradus stars defined by \cite{Dupret2005} (Figure~\ref{Fig1_strip}).
\cite{Rimoldini2012} classified HIP\,106223 as a $\gamma$\,Doradus variable with  probabilities of 0.27 (RF method) and 0.63 (MB method). However, \cite{Rimoldini2012} did not find low frequency pulsations in this star, a dominant signal was derived at 12.79676\,$c/d$.

We observed this star during 28 runs. We found that HIP\,106223 is different from other candidates to $\delta$\,Scuti stars. It has obvious low frequency pulsations and no pulsations at higher frequencies (Figure~\ref{Fig13_FT_106223} and Table~\ref{table:2}). 
 Most of the time there were observed just slow increases or decreases of the magnitude during a night (Figure~\ref{Fig14_LC_106223}). Brightness of the star increased more significantly on JD\,2457625.4, it reached the maximum and started to decrease the same night.
 The amplitude spectrum of the combined LC gave a dominant frequency at 1.1429\,$c/d$ with amplitude of 31.09\,mmag. After prewhitening of this frequency it was still possible to prewhiten two more frequencies (1.1165\,$c/d$ and 1.3572\,$c/d$ with amplitudes 16.13\,mmag and  11.03\,mmag, respectively). As this region of low frequencies may be affected by instrumental or weather condition instabilities, the two signals of lower amplitudes still have to be confirmed. Except these three low frequencies there were no more significant peaks in a full range of the amplitude spectrum. It was only a small peak at 14.0649\,$c/d$ with ${\rm S/N}=3.05$, which is close to the one derived by \cite{Rimoldini2012}.

 However, the LC calculated using only these three low frequencies can not reach the maximal brightness of HIP\,106223 observed on JD 2457625.4. There could be more than three frequencies of pulsations which sometimes interfere creating a highly increased amplitude of pulsations. We evaluated that the interference of 10 frequencies determined after prewhitening of the amplitude spectrum can give a satisfactory match of the observed and calculated LCs. 
 
 Summarizing all the information collected about HIP\,106223, most probably it is a $\gamma$\,Doradus star or $\delta$\,Scuti--$\gamma$\,Doradus hybrid star with very low amplitude pulsations at higher frequencies.
 
\subsection{HIP\,107786}

   \begin{figure}[!h]
   \centering
   \includegraphics[width=\hsize]{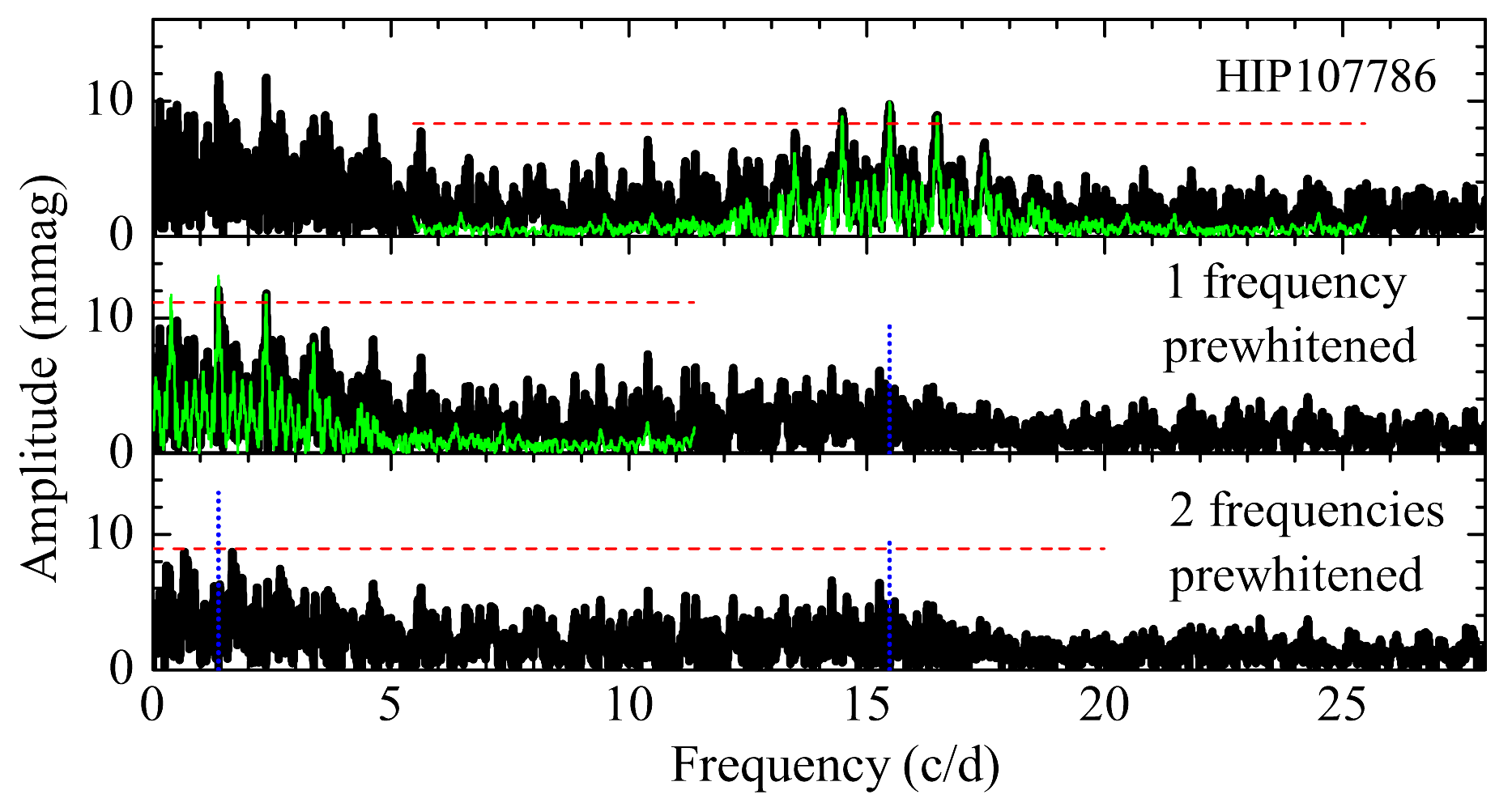}
      \caption{Amplitude spectrum of the combined LC of HIP\,107786 and its prewhitening. Meanings of the lines are the same as in Figure~\ref{Fig5_FT_2923}.  }  
         \label{Fig15_FT_107786}
   \end{figure}

HIP\,107786 is the most luminous star among our targets, moreover, it is a triple system with the $\delta$\,Scuti type star (\citealt{Henry2004}, \citealt{Fekel2015}). The short orbital period of the Aa and Ab pair is 1.4707386 days, and the long period of A and B components is 724.09 days. According to fundamental parameters derived by \cite{McDonald2012}, this system is located inside the overlapping instability strips of $\delta$\,Scuti and $\gamma$\,Doradus stars (\citealt{Xiong2016}), but it is out of the $\gamma$\,Doradus instability strip derived by \cite{Dupret2005}.
Different components of the triple system are located at different places in the HR diagram. According to \citet{Fekel2015}, this system consists of a broad-lined A8\,V star, which is a variable $\delta$\,Scuti type star, and an unseen mid-M dwarf companion (Aa and Ab components). One more component (B) is F7\,V star at a larger distance from the close pair of Aa and Ab. A total estimated mass of Aa plus Ab is 2.1 ${{M}_{\rm{Sun}}}$, and 1.2 ${{M}_{\rm{Sun}}}$ of the component B. The mass of $\delta$\,Scuti star (Aa) and its companion Ab may be equal to 1.9 and 0.2~${{M}_{\rm{Sun}}}$, respectively \citep{Fekel2015}. That means that the $\delta$\,Scuti companion of HIP\,107786 triple system should be located at slightly lower luminosities and temperatures than it is showed in the HR diagram (Figure~\ref{Fig1_strip}).

\cite{Henry2004} found three frequencies of photometric brightness variations in HIP\,107786 (1.35980\,$c/d$, 15.43448\,$c/d$ and 15.78034\,$c/d$) with amplitudes between 30\,mmag and 11\,mmag. Two of the derived frequencies are typical for $\delta$\,Scuti stars. But the lowest one was explained by \cite{Henry2004} as a result from the ellipticity effect of Aa component. 

We had 14 runs of observations for the field with HIP\,107786. The amplitude spectrum of these observations gave clear signals at 15.4817\,$c/d$ with amplitude of 9.87~mmag and at 1.3778\,$c/d$ with amplitude of 13.11\,mmag (Figure~\ref{Fig15_FT_107786} and Table~\ref{table:2}). These two frequencies fit well to the frequencies derived by \cite{Henry2004}, however we did not find 
the second signal between 15\,$c/d$ and 16\,$c/d$.

\subsection{HIP\,113487}

   \begin{figure}[!h]
   \centering
   \includegraphics[width=\hsize]{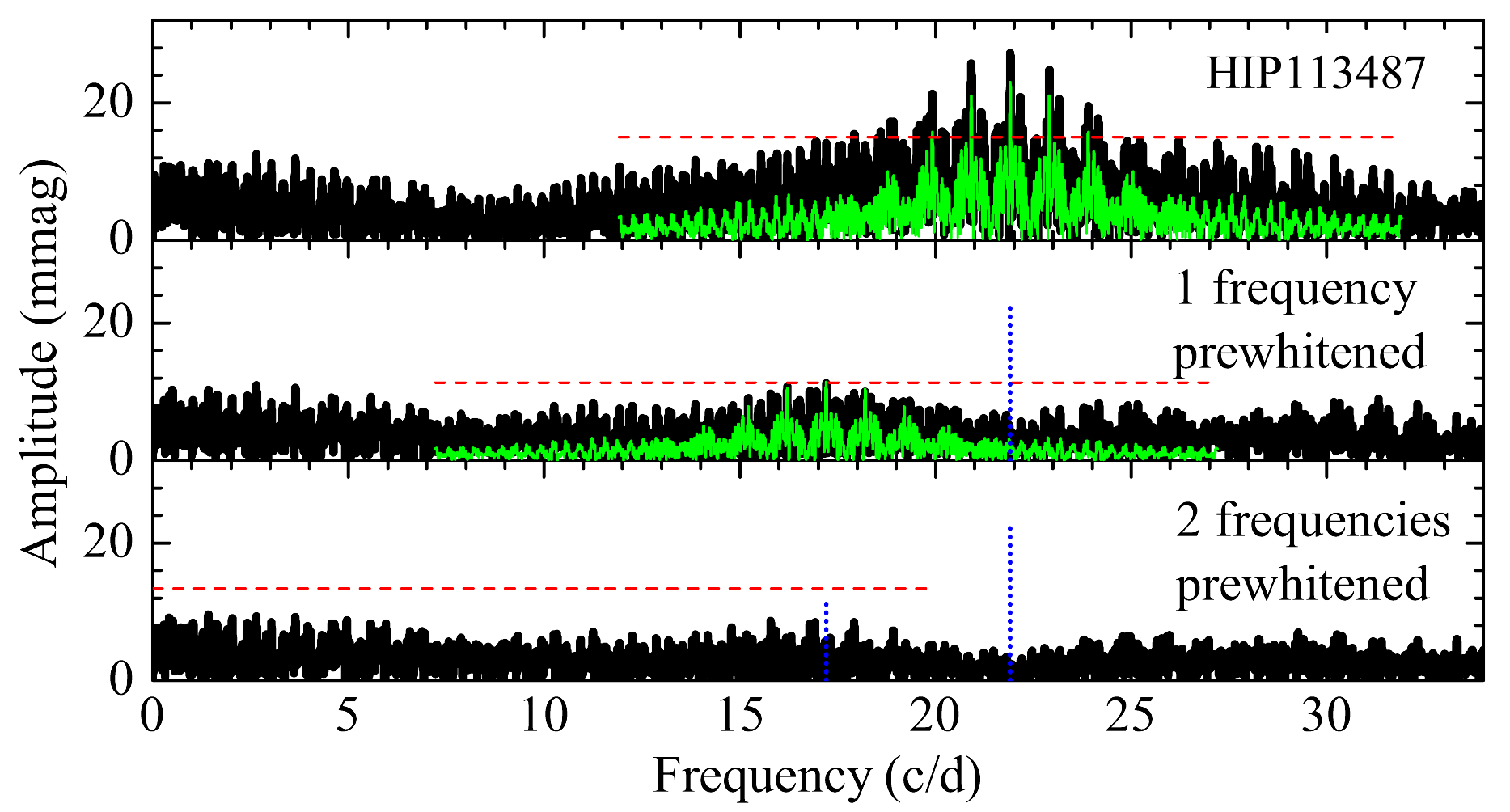}
      \caption{Amplitude spectrum of the combined LC of HIP\,113487 and its prewhitening. Meanings of the lines are the same as in Figure~\ref{Fig5_FT_2923}.  }  
         \label{Fig16_FT_113487}
   \end{figure}

According to fundamental parameters derived by \cite{McDonald2012}, it appears on the blue edge of instability strip of $\gamma$\,Doradus stars derived by \cite{Xiong2016} (Figure~\ref{Fig1_strip}). 
\cite{Rimoldini2012} classified HIP\,113487 as a low amplitude $\delta$\,Scuti variable with probabilities of 0.41 (RF method) and 0.58 (MB method). \cite{Rimoldini2012} also derived  frequency of the dominant signal, which was 20.91221\,$c/d$.

HIP\,113487 had the highest frequency of pulsations among the observed stars, amplitude of the dominant signal was high and easily determined. The amplitude spectrum analysis provided a strong signal at 21.9102\,$c/d$ with amplitude of 23.01\,mmag and ${\rm S/N}=6.16$, and one more signal at 17.2064\,$c/d$ with amplitude of 11.40\,mmag and ${\rm S/N}=4.04$. Both frequencies are typical for $\delta$\,Scuti stars (Figure~\ref{Fig16_FT_113487} and Table~\ref{table:2}). 

\subsection{HIP\,115093}

   \begin{figure}[!h]
   \centering
   \includegraphics[width=\hsize]{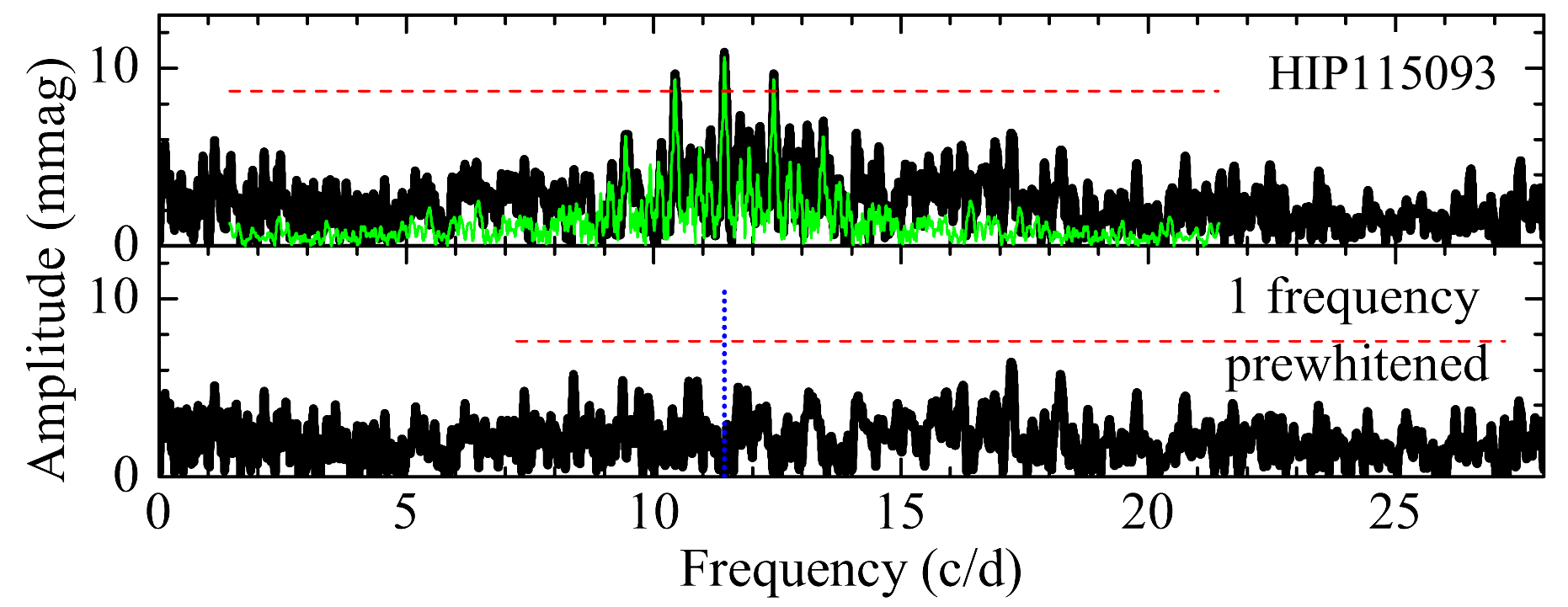}
      \caption{Amplitude spectrum of the combined LC of HIP\,115093 and its prewhitening. Meanings of the lines are the same as in Figure~\ref{Fig5_FT_2923}.  }  
         \label{Fig17_FT_115093}
   \end{figure}

According to fundamental parameters derived by \cite{McDonald2012}, HIP\,115093 appears in the HD diagram inside the overlapping instability strips of $\delta$\,Scuti and $\gamma$\,Doradus stars (\citealt{Xiong2016}) (Figure~\ref{Fig1_strip}). 
\cite{Handler1999} used Hipparcos observations and classified HIP\,115093 as an object, which may be a $\gamma$\,Doradus star with the frequency of 0.5587\,$c/d$, but whose nature remained uncertain. Latter \cite{Handler2002} classified it as a $\delta$\,Scuti candidate. 
\cite{Rimoldini2012} classified HIP\,115093 as a low amplitude $\delta$\,Scuti variable with probabilities of 0.53 (RF method) and 0.49 (MB method). \cite{Rimoldini2012} also derived  frequency of the dominant signal, which was 10.27306\,$c/d$.

We had 10 runs of observations for the field of this star. The amplitude spectrum of observed LC gave a single signal with ${\rm S/N}=4.87$ at 11.4318\,$c/d$ with amplitude of 10.61~mmag (Figure~\ref{Fig18_FT_115856} and Table~\ref{table:2}), which is intrinsic to $\delta$\,Scuti type stars. This star might have more modes of pulsations  hidden in a quite high noise level of its amplitude spectrum. If there is any pulsation typical to $\gamma$\,Doradus stars, it should have a small amplitude and we were not able to detect it.

\subsection{HIP\,115856}

   \begin{figure}[!h]
   \centering
   \includegraphics[width=\hsize]{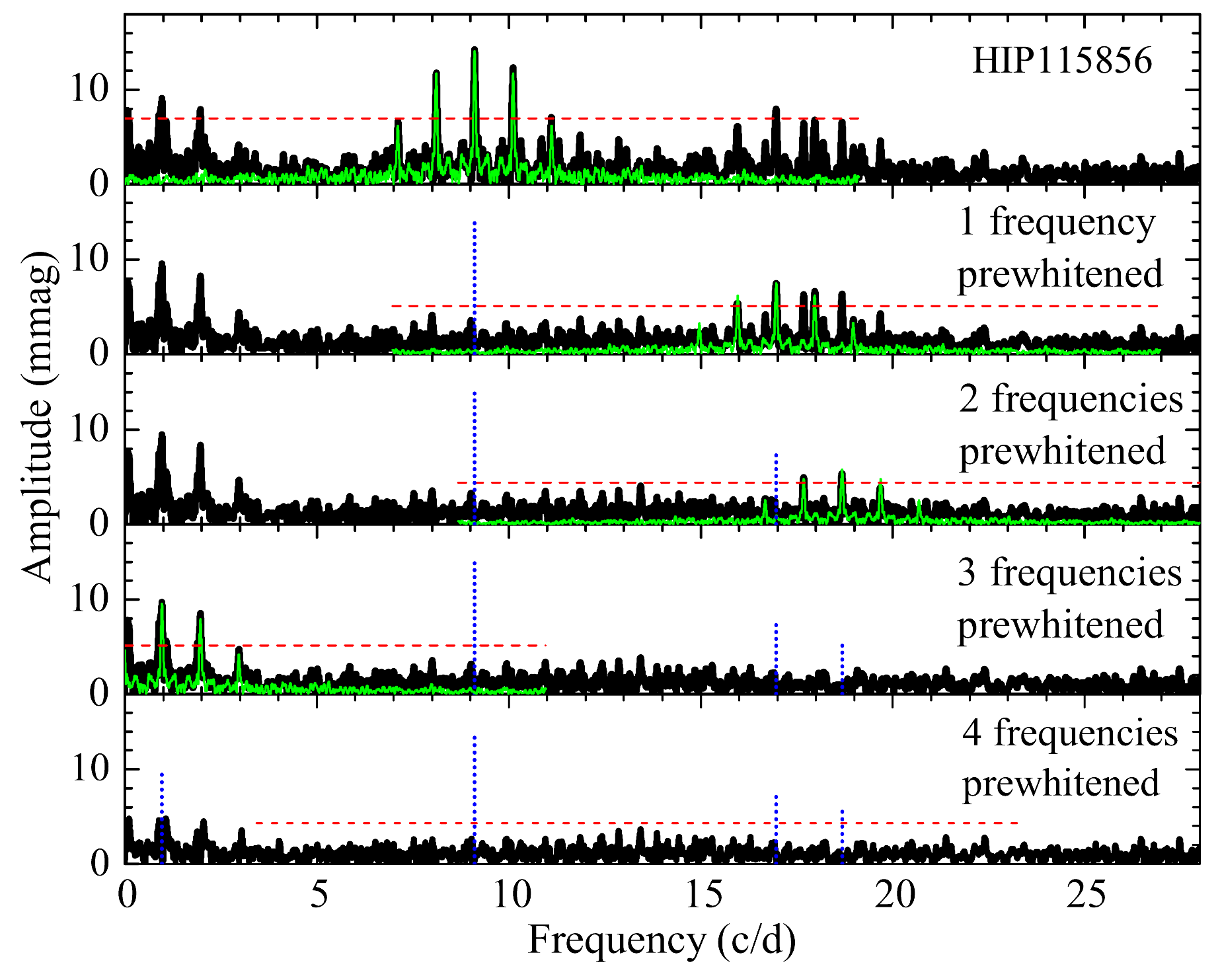}
      \caption{Amplitude spectrum of the combined LC of HIP\,115856 and its prewhitening process. Meanings of the lines are the same as in Figure~\ref{Fig5_FT_2923}.  
      }  
         \label{Fig18_FT_115856}
   \end{figure}

According to fundamental parameters derived by \cite{McDonald2012}, HIP\,115856 appears in the HD diagram inside the overlapping instability strips of $\delta$\,Scuti and $\gamma$\,Doradus stars (\citealt{Xiong2016}) (Figure~\ref{Fig1_strip}). Positions of HIP\,115856 and HIP\,115093 are  very close, however, the amplitude spectrum of HIP\,115856 is much richer.
\cite{Rimoldini2012} classified HIP\,115093 as a low amplitude $\delta$\,Scuti variable with  probabilities of 0.69 (RF method) and 0.74 (MB method). \cite{Rimoldini2012} also derived frequency of the dominant signal, which was 9.10961\,$c/d$.

Seventeen runs of HIP\,115856 observations gave an amplitude spectrum with at least 4 signals, with the strongest one at 9.1109\,$c/d$ and amplitude of 14.08~mmag (Figure~\ref{Fig18_FT_115856} and Table~\ref{table:2}). 
We found at least one low frequency at  0.9669\,$c/d$ with amplitude of 9.55\,mmag and ${\rm S/N}=7.43$. 
Thus we conclude that HIP\,115856 expose typical brightness variations of $\delta$\,Scuti stars, however there is a possibility that it may be a $\delta$\,Scuti--$\gamma$\,Doradus hybrid star.

\section{Conclusions}

We obtained 24\,215 CCD images and analyzed stellar light curves of thirteen $\delta$\,Scuti candidates selected from the Hipparcos catalog.  
We confirm that twelve of them are variables and pulsate with frequencies typical to $\delta$\,Scutti type stars. 
Moreover, five of them (HIP\,2923, HIP\,5526, HIP\,11090, HIP\,115856, and HIP\,106219) may be hybrid $\delta$\,Scuti-$\gamma$\,Doradus pulsators, as they simultaneously show high-frequency pulsations typical to the $\delta$\,Scuti stars and significant low-frequency oscillations (between 0.5422\,$c/d$ and 1.3778\,$c/d$) characteristic to the $\gamma$\,Doradus stars.
One more star, HIP\,106223, pulsate just with low frequencies typical to variables of $\gamma$\,Doradus type stars.

\section*{Acknowledgements}

This research has made use of the SIMBAD database and NASA's Astrophysics Data System (operated at CDS, Strasbourg, France), and was funded by the grant from the Research Council of Lithuania (LAT-08/2016). We are grateful to an anonymous reviewer for providing insightful comments and
directions for additional analysis of observations which have resulted in this paper.




\end{document}